\documentclass[amsmath,aps,showpacs,a4paper,10pt]{revtex4}

 \usepackage{epsf}
 \usepackage{graphicx}    

 \usepackage{rotating}    

\begin{document}

 \newcommand{\be}[1]{\begin{equation}\label{#1}}
 \newcommand{\ee}{\end{equation}}
 \newcommand{\bea}{\begin{eqnarray}}
 \newcommand{\eea}{\end{eqnarray}}
 \def\disp{\displaystyle}

 \begin{titlepage}

 \begin{flushright}
 arXiv:2504.13380
 \end{flushright}

 \title{\Large \bf Alleviating the Hubble Tension with a Local Void\\
 and Transitions of the Absolute Magnitude}

 \author{Jing-Yi~Jia\,$^{a,}$\footnote{email
 address:\ jjy@bit.edu.cn}\,,
 Jia-Lei~Niu\,$^a$\,,
 Da-Chun~Qiang\,$^b$\,,
 Hao~Wei\,$^{a,}$\footnote{Corresponding author;\ email
 address:\ haowei@bit.edu.cn}\vspace{2.4mm}}
 \affiliation{$^{a)\,}$School of Physics, Beijing
 Institute of Technology, Beijing 100081, China\vspace{2mm}\\
 $^{b)\,}$Institute for Gravitational Wave Astronomy, Henan Academy of
 Sciences, Zhengzhou 450046, Henan, China}

 \begin{abstract}\vspace{1cm}
 \centerline{\bf ABSTRACT}\vspace{2mm}
 Nowadays, one of the well-known serious challenges in cosmology is the
 Hubble tension, namely the discrepancy between the Hubble constants
 from the local observation of Type Ia supernova (SNIa) and the high-$z$
 observation of cosmic microwave background (CMB). Here, we are
 interested in alleviating the Hubble tension with a local void. The key
 idea is assuming that we live in a locally underdense void, where one
 will feel a faster expansion rate compared to the cosmic average.~In
 the literature, it was found that a local void cannot satisfyingly
 alleviate the Hubble tension, since it is not preferred over
 the $\Lambda$CDM model by the observations such as the Pantheon SNIa
 sample, especially in terms of the information criteria AIC and BIC.~In
 the present work, we try to alleviate the Hubble tension with a local
 void and transitions of the absolute magnitude $M$, by using
 the Pantheon+ SNIa sample alone or jointly with the CMB data of Planck
 2018.~We find that the Hubble tension can be satisfyingly alleviated,
 while the $\Lambda$LTB void models are strongly preferred by
 the observations.
 \end{abstract}

 \pacs{98.80.Es, 98.65.Dx, 98.80.-k}

 \maketitle

 \end{titlepage}

 \renewcommand{\baselinestretch}{1.0}


\section{Introduction}\label{sec1}

Nowadays, one of the well-known serious challenges in cosmology is the
 Hubble tension~\cite{DiValentino:2025sru,Abdalla:2022yfr,
 Perivolaropoulos:2021jda,Verde:2023lmm,DiValentino:2022fjm,
 Efstathiou:2024dvn,DiValentino:2021izs,Rong-Gen:2023dcz,Hu:2023jqc,
 Chang:2022tzj}.~That~is, there is a significant tension between the
 Hubble constants $H_0$ measured from various independent probes in
 the early and late/local universes.~In particular, assuming
 the well-known $\Lambda$CDM model, the Hubble constant inferred from
 the final full-mission Planck measurements (Planck 2018) of
 the cosmic microwave background (CMB) is given
 by $H_0=67.36\pm 0.54\;{\rm km/s/Mpc}$~\cite{Planck:2018vyg}.~Note
 that it cannot be substantially changed by assuming various trivial
 extensions to the base-$\Lambda$CDM model.~On the other hand, based
 on the Cepheid/Type Ia supernova (SNIa) distance ladder, the local
 determination of $H_0$ from the Hubble Space Telescope (HST)
 and the SH0ES team (R22) is given by $H_0=73.04\pm 1.04\;{\rm
 km/s/Mpc}$~\cite{Riess:2021jrx}.~Obviously, this Hubble
 constant directly from the local measurement of Cepheids/SNIa is in
 $>5\sigma$ tension with the one indirectly inferred from CMB in the
 early universe (at high redshift $z_\ast\sim 1090$).~Note that
 the recent SH0ES 2024 result (R24) $H_0=73.17\pm 0.86\;{\rm
 km/s/Mpc}$~\cite{Riess:2024vfa} is still in $5-6\sigma$ tension with
 Planck 2018, while it has been cross-checked with the early James Webb
 Space Telescope (JWST) subsamples from SH0ES and CCHP.

In history, the Hubble tension has not emerged in the WMAP era of CMB
 before 2011, then manifested itself at $2-3\sigma$ since the first
 Planck result in 2013, and finally reached $>5\sigma$ in December 2021
 (R22). Various independent probes, e.g.~the Tip of the Red Giant Branch
 (TRGB), Mira variables, J-region Asymptotic Giant Branch (JAGB) stars,
 baryon acoustic oscillations (BAO), strong gravitational
 lensing, gravitational waves, water masers, have been considered in the
 past decade to this end, but the Hubble tension has not been resolved
 to date.~We refer to e.g.~\cite{DiValentino:2025sru,Abdalla:2022yfr,
 Perivolaropoulos:2021jda,Verde:2023lmm,DiValentino:2022fjm,
 Efstathiou:2024dvn,DiValentino:2021izs,Rong-Gen:2023dcz,Hu:2023jqc,
 Chang:2022tzj} for comprehensive reviews.

Many efforts have been made in the literature.~Of course, it is fairly
 reasonable to carefully check the observational data, since there might
 be unresolved systematic errors in them.~On the other hand, if all the
 observational data are right, the Hubble tension suggests the need for
 non-trivial extensions to the standard $\Lambda$CDM cosmology. There
 might be something new in the early, middle, late or local universes.
 Although numerous theoretical solutions have been proposed
 in the literature (e.g.~\cite{DiValentino:2025sru,Abdalla:2022yfr,
 Perivolaropoulos:2021jda,Verde:2023lmm,DiValentino:2022fjm,
 Efstathiou:2024dvn,DiValentino:2021izs,Rong-Gen:2023dcz,
 Hu:2023jqc,Chang:2022tzj,Montani:2024pou,Montani:2024ntj,
 Dainotti:2021pqg,Schiavone:2022wvq,Silva:2023wvs,Vagnozzi:2023nrq} and
 references therein), the Hubble tension is still a tough problem currently.

Here, we are interested in alleviating the Hubble tension with
 a local void.~The key idea is to violate the cosmological principle by
 assuming that we live in a locally underdense void centered nearby our
 location.~In this underdense region, one will feel a locally faster
 expansion rate compared to the cosmic average.~Noting that $H_0\sim
 73\;{\rm km/s/Mpc}$ directly measured from the local Cepheids/SNIa is
 larger than $H_0\sim 67\;{\rm km/s/Mpc}$ indirectly inferred from CMB
 assuming a background Friedmann-Robertson-Walker (FRW) universe, the
 Hubble tension might be alleviated with a local
 void~\cite{DiValentino:2025sru,Abdalla:2022yfr,DiValentino:2021izs,
 Rong-Gen:2023dcz}.

A local void can be approximately modeled by using the
 Lema\^{i}tre-Tolman-Bondi (LTB) metric~\cite{Lemaitre:1933gd,
 Tolman:1934za,Bondi:1947fta} firstly proposed in 1933, which
 is spherically symmetric and radially inhomogeneous.~In 1998,
 the nearby sample has been examined for evidence of a local
 ``\,Hubble Bubble\,''~\cite{Zehavi:1998gz}.~Later, various LTB void
 models were used to explain the cosmic acceleration without invoking
 dark energy or modified gravity~\cite{Celerier:1999hp,Celerier:2007jc,
 Barrett:1999fd,Tomita:2000jj,Tomita:2001gh,Iguchi:2001sq,
 Zhang:2012qr,Wang:2011kj,Enqvist:2006cg,Enqvist:2007vb,
 GarciaBellido:2008nz,Garcia-Bellido:2008sdt,Celerier:2011zh,
 Celerier:2012xr,Alnes:2005rw,Celerier:2009sv,
 Vanderveld:2006rb,Zibin:2008vj,Zibin:2011ptm,Yan:2014eca,Yu:2019cku,
 Alnes:2006pf,Sundell:2015cza,ChirinosIsidro:2016vah}.
 Notice that cosmological constant is not allowed to this end, since it
 is the simplest candidate of dark energy as well known.~Recently, the
 Hubble tension attracted a lot of attention.~If the goal is no longer
 to explain the cosmic acceleration, we can allow a non-zero
 cosmological constant $\Lambda$ in the LTB model and use
 the $\Lambda$LTB void model to alleviate the Hubble tension, while
 the cosmic acceleration is instead driven by $\Lambda\not=0$.~The role
 of LTB void has been changed in this case.

In 2013, the evidence for a $\sim 300\;{\rm Mpc}$ local void
 was claimed by Keenan, Barger and Cowie (KBC)~\cite{Keenan:2013mfa}.
 In 2018, Hoscheit and Barger~\cite{Hoscheit:2018nfl} claimed that the
 KBC void can significantly reduce the Hubble tension in the
 $\Lambda$LTB framework.~Later in 2018, Shanks~{\it et
 al.}~\cite{Shanks:2018rka} also claimed that a local void can
 satisfyingly relieve the Hubble tension.~However, both results
 in~\cite{Hoscheit:2018nfl,Shanks:2018rka} were quickly criticized in
 2019 by Kenworthy, Scolnic and Riess~\cite{Kenworthy:2019qwq}.~But the
 debate was not settled in fact.~Lukovi\'c~{\it
 et al.}~\cite{Lukovic:2019ryg} continued the discussions in 2019, and
 Kazantzidis and Perivolaropoulos~\cite{Kazantzidis:2020tko} claimed
 hints of a local void in 2020.~At the end of 2020, Cai~{\it et
 al.}~\cite{Cai:2020tpy} claimed again that the Hubble tension cannot
 be saved by a local void alone.

It is worth noting that the SNIa sample used
 in~\cite{Hoscheit:2018nfl,Kenworthy:2019qwq,Lukovic:2019ryg,
 Kazantzidis:2020tko,Cai:2020tpy} mentioned above is the
 Pantheon sample~\cite{Pan-STARRS1:2017jku} released in 2017.~But as is
 well known, the Hubble constant $H_0$ is heavily degenerated with the
 absolute magnitude $M$ of SNIa in the Pantheon sample (and other SNIa
 samples before Pantheon).~Commonly, they can be marginalized
 as a combination ${\cal M}=M+5\log\left(c/H_0/{\rm Mpc}\right)+25$ when
 we constrain other cosmological parameters, where $c$ is the speed of
 light and ``\,$\log$\,'' gives the logarithm to base $10$.~This is the
 reason to say ``\,SNIa samples are Hubble-free\,'' in the past.~Thus,
 in order to infer the Hubble constant $H_0$ from the Pantheon
 SNIa sample, the absolute magnitude $M$ should be given {\it a priori}
 (usually the one from SH0ES is adopted). In the works of local
 ($\Lambda$LTB) void mentioned above, the absolute magnitude~$M$ of SNIa
 has also been given {\it a priori}, or $\cal M$ was fitted to data as a
 free parameter alternatively.

The situation has been changed since the Pantheon+ SNIa
 sample~\cite{Brout:2022vxf,Scolnic:2021amr,PantheonPlusSH0ES} released
 in 2022.~In the Pantheon+ sample, the Cepheid calibrated host-galaxy
 distance moduli are also provided by SH0ES, which can be
 used to constrain the absolute magnitude $M$, and hence the degeneracy
 between $H_0$ and $M$ are broken.~So, $H_0$ and $M$ can be separately
 constrained as free parameters.~This makes the Pantheon+ sample much
 better than the Pantheon sample.~On the other hand, the new Pantheon+
 sample consists of~more SNIa (especially more SNIa at low redshifts)
 than the old Pantheon sample.~Thus, it is reasonable to study
 the Hubble tension with the Pantheon+ SNIa sample instead.

Interestingly, a bulk flow around $z\sim 0.1$ and a local infall around
 $z\sim 0.04$ were claimed by using the Pantheon+ SNIa sample
 in~\cite{Sorrenti:2022zat,Sorrenti:2024ztg,
 Sorrenti:2024ugq}.~Similarly, the bulk flow in the local universe was
 also found with the Pantheon+ SNIa sample in~\cite{Lopes:2024vfz} and
 with CosmicFlows4 in~\cite{Watkins:2023rll}.~On the other hand, a local
 inhomogeneity was claimed in~\cite{Sanejouand:2023jkv} with
 the Pantheon+ SNIa sample.~In~\cite{Cai:2021wgv}, a late-time inhomogeneous
 resolution for the Hubble tension with a chameleon dark energy was
 proposed.~So, well motivated by these works, it is still interesting to
 consider a local inhomogeneity (especially a local void)
 for the Hubble tension in the new era of the Pantheon+ SNIa sample.

Usually, the absolute magnitude $M$ of SNIa is assumed to be a universal
 constant, or two-value constants $M=M_1$ for host stellar mass
 $<10^{10}M_\odot$ and $M=M_2$ otherwise~\cite{Pan-STARRS1:2017jku,
 SDSS:2014iwm,Wang:2015tua}, where $M_\odot$ is the solar mass.~But this
 does not affect the cosmological constraints, since $M$ will
 be marginalized with $H_0$ as a combination ${\cal M}=M+5\log\left(
 c/H_0/{\rm Mpc}\right)+25$ (see Appendix~C of~\cite{SNLS:2011lii} for
 details).~These are standard procedures for SNIa samples like Pantheon
 and before.~However, the Pantheon+ sample is different, as mentioned
 above, in which $H_0$ and $M$ are separately constrained as
 free parameters.~In~\cite{Perivolaropoulos:2023iqj}, it is found that
 the absolute magnitude $M$ in the Pantheon+ sample has a transition
 at the distance $d_{\rm crit}\sim 20\;{\rm Mpc}$, namely $M= M_<$ for
 $d<d_{\rm crit}$ and $M= M_>$ for $d>d_{\rm crit}$, while
 the evidence is convincing in terms of $\Delta\chi^2$ and the Akaike
 information criterion (AIC).~Notice that this transition of $M$ does
 not affect the constraints on other cosmological parameters
 e.g.~$\Omega_m$, but slightly raises the best-fit $H_0$, and hence
 the Hubble tension cannot be
 addressed~\cite{Perivolaropoulos:2023iqj}.~Recently, two transition
 models for the absolute magnitude $M$ in the Pantheon+ sample were
 considered in~\cite{Liu:2024vlt}, namely (a) sudden transition: $M=M_0$
 if $z<z_t$ and $M=M_0+A$ if $z\geq z_t$, (b) linear transition: $M(z)=M_0$,
 $M_0+A\left(z-z_0\right)/\left(z_t-z_0\right)$, $M_0+A$ if $z<z_0$,
 $z_0\leq z<z_t$, $z\geq z_t$, respectively.~Notice that the transitions are
 given in terms of redshift $z_t$, rather than distance $d_{\rm crit}$ as in
 \cite{Perivolaropoulos:2023iqj}.~It was claimed in~\cite{Liu:2024vlt}
 that the Hubble tension could be alleviated by the sudden transition model
 with a mild evidence in terms of AIC (while the linear transition model
 failed in terms of AIC), but both the sudden and linear transition
 models failed in terms of the Bayesian information criterion (BIC).

Note that both~\cite{Perivolaropoulos:2023iqj,Liu:2024vlt} are in the
 framework of $\Lambda$CDM cosmology, and they cannot alleviate the
 Hubble tension in terms of BIC (even in terms of AIC, the evidence for
 the sudden transition model in~\cite{Liu:2024vlt} is just ``\,mild\,'',
 but not ``\,strong\,'').~Well motivated by the works mentioned above,
 we try to alleviate the Hubble tension with a local $\Lambda$LTB void
 and transitions of the absolute magnitude $M$ in the present work. The
 evidence will be strong simultaneously in terms of $\Delta\chi^2$,
 AIC, BIC and Bayesian evidence.

The rest of this paper is organized as follows.~In Sec.~\ref{sec2}, we
 briefly introduce the $\Lambda$LTB void model, which will be compared
 with the fiducial models in terms of the information criteria AIC and
 BIC. In Secs.~\ref{sec3} and \ref{sec4}, we test the models with the
 Pantheon+ SNIa sample, and SNIa plus CMB, respectively. We will see
 whether the Hubble tension can be alleviated by using the $\Lambda$LTB
 void models and transitions of the absolute magnitude $M$.~Finally,
 a brief conclusion and some discussions are given in Sec.~\ref{sec5}.


\section{The local $\Lambda$LTB void and the preliminary}\label{sec2}


\subsection{The local $\Lambda$LTB void}\label{sec2a}

At first, we briefly introduce the $\Lambda$LTB void model, in which
 the universe is spherically symmetric and radially inhomogeneous, and
 we are living in a locally underdense void centered nearby our
 location. In comoving coordinates ($r$, $\theta$, $\phi$) and
 synchronous time $t$, the LTB metric is given by~\cite{Lemaitre:1933gd,
 Tolman:1934za,Bondi:1947fta} (see also e.g.~\cite{Zhang:2012qr,Wang:2011kj,
 Enqvist:2006cg,Enqvist:2007vb,GarciaBellido:2008nz,Garcia-Bellido:2008sdt,
 Celerier:2011zh,Celerier:2012xr,Alnes:2005rw,Yan:2014eca,
 Hoscheit:2018nfl,Kenworthy:2019qwq,Cai:2020tpy})
 \vspace{-10mm} 
 \be{eq1}
 ds^2=c^2 dt^2-\frac{R^{\prime\,2}(r,t)}{1-k(r)}\,dr^2-
 R^2(r,t)\,d\Omega^2\,,
 \ee
 where $c$ is the speed of light, $d\Omega^2=d\theta^2+\sin^2\theta\,
 d\phi^2$, a prime denotes a derivative with respect to $r$, and $k(r)$
 is an arbitrary function of $r$ playing the role of spatial
 curvature.~Notice that it reduces to the well-known FRW metric
 if $R(r,t)=a(t)\,r$ and $k(r)=kr^2$.~Considering the universe filled by
 dust matter and a non-zero cosmological constant $\Lambda$
 (equivalently the vacuum energy), the Friedmann equation
 for the $\Lambda$LTB model reads~\cite{Enqvist:2006cg,Enqvist:2007vb,
 Celerier:2011zh,Celerier:2012xr,Hoscheit:2018nfl}
 \be{eq2}
 \frac{\dot{R}^2}{c^2}=-k(r)+\frac{2m(r)}{R}+\frac{\Lambda}{3}R^2\,,
 \ee
 where a dot denotes a derivative with respect to $t$. In fact,
 Eq.~(\ref{eq2}) is a first integral of the second Einstein
 field equation, and $m(r)$ is arbitrary (non-negative) function of
 $r$ from the integral, playing the role of gravitational mass within
 the comoving spherical shell. Introducing the Hubble parameter
 \be{eq3}
 H(r,t)\equiv\frac{\dot{R}(r,t)}{R(r,t)}\,,
 \ee
 and the fractional energy densities at the present time $t_0$,
 \be{eq4}
 \Omega_m(r)\equiv\frac{2m(r)\,c^2}{H_0^2(r)\,R_0^3(r)}\,,\quad
 \Omega_k(r)\equiv\frac{-k(r)\,c^2}{H_0^2(r)\,R_0^2(r)}\,,\quad
 \Omega_\Lambda(r)\equiv\frac{\Lambda c^2}{3H_0^2(r)}\,,
 \ee
 where the subscript ``\,0\,'' indicates the value of corresponding
 quantity at the present time $t_0$, Eq.~(\ref{eq2}) evaluated
 at the present time $t_0$ gives
 \be{eq5}
 1=\Omega_m(r)+\Omega_k(r)+\Omega_\Lambda(r)\,.
 \ee
 With Eq.~(\ref{eq4}), we can recast Eq.~(\ref{eq2}) as
 \be{eq6}
 H^2(r,t)=\left[\frac{\dot{R}(r,t)}{R(r,t)}\right]^2=H_0^2(r)
 \left[\,\Omega_m(r)\,\frac{R_0^3(r)}{R^3(r,t)}+\Omega_k(r)\,
 \frac{R_0^2(r,t)}{R^2(r,t)}+\Omega_\Lambda(r)\,\right]\,.
 \ee
 Choosing the conventional gauge $R(r,t_0)=R_0(r)=r$ (see
 e.g.~\cite{Zhang:2012qr,Wang:2011kj,Enqvist:2006cg,Enqvist:2007vb,
 GarciaBellido:2008nz,Garcia-Bellido:2008sdt,Celerier:2011zh,
 Celerier:2012xr,Alnes:2005rw,Yan:2014eca,
 Hoscheit:2018nfl,Kenworthy:2019qwq,Cai:2020tpy}), it becomes
 \be{eq7}
 \dot{R}(r,t)=R(r,t)\,H_0(r)\left[\,\Omega_m(r)\,\frac{r^3}{R^3(r,t)}
 +\Omega_k(r)\,\frac{r^2}{R^2(r,t)}+\Omega_\Lambda(r)\,\right]^{1/2}\,.
 \ee
 Note that Eq.~(\ref{eq7}) can be integrated and then inverted to solve
 for $R(r,t)$~\cite{Kenworthy:2019qwq}.~In order to compare the
 theoretical LTB model with observations, we need to associate
 the coordinates with redshift $z$.~For an observer located at
 the center $r=0$, by symmetry, incoming light travels along
 radial null geodesics, $ds^2=d\Omega^2=0$, and hence we have
 \cite{Enqvist:2006cg,Enqvist:2007vb,GarciaBellido:2008nz,
 Garcia-Bellido:2008sdt,Yan:2014eca,
 Hoscheit:2018nfl,Kenworthy:2019qwq,Cai:2020tpy}
 \be{eq8}
 \frac{dt}{dr}=-\frac{R^\prime (r,t)}{c\sqrt{1-k(r)}}\,.
 \ee
 On the other hand, the relation of the radial variable $r$ to
 redshift $z$ is given by~\cite{Enqvist:2006cg,Enqvist:2007vb,
 GarciaBellido:2008nz,Garcia-Bellido:2008sdt,Yan:2014eca,
 Hoscheit:2018nfl,Kenworthy:2019qwq,Cai:2020tpy}
 \be{eq9}
 \frac{1}{1+z}\,\frac{dz}{dr}=\frac{\dot{R}^\prime (r,t)}{c\sqrt{1-k(r)}}\,.
 \ee
 Eqs.~(\ref{eq8}) and (\ref{eq9}) can be numerically integrated to find
 the cosmic time $t$ and the radial coordinate $r$ as functions
 of redshift $z$~\cite{Kenworthy:2019qwq}.~Finally, the
 luminosity distance is given by~\cite{Enqvist:2006cg,Enqvist:2007vb,
 GarciaBellido:2008nz,Garcia-Bellido:2008sdt,Yan:2014eca,
 Hoscheit:2018nfl,Kenworthy:2019qwq,Cai:2020tpy}
 \be{eq10}
 d_L(z)=(1+z)^2 R(r(z),t(z))\,.
 \ee
 To numerically solve the differential equations
 $(\ref{eq7})-(\ref{eq9})$, the fractional densities $\Omega_m(r)$,
 $\Omega_k(r)$, $\Omega_\Lambda(r)$ and their boundary conditions, as
 well as the boundary condition on $H_0(r)$, are
 needed~\cite{Kenworthy:2019qwq}.

Here, we consider a spherically symmetric local void surrounded
 by a flat $\Lambda$CDM background FRW universe.~In this work, we use
 the (constrained) Garcia-Bellido-Haugboelle (GBH) profile function
 \cite{GarciaBellido:2008nz,Garcia-Bellido:2008sdt} (see also
 e.g.~\cite{Yan:2014eca,Hoscheit:2018nfl,Kenworthy:2019qwq,Cai:2020tpy}) to
 model the mass distribution of the void, namely
 \be{eq11}
 \delta(r)=\delta_V\left[\,\frac{1-\tanh\left((r-r_V)/2\Delta_r\right)}
 {1+\tanh\left(r_V/2\Delta_r\right)}\,\right]\,,
 \ee
 with void depth $\delta_V$, void characteristic radius $r_V$,
 and transition width $\Delta_r$ to uniformity. The fractional deficit
 $\delta(r)$ is given by $\delta(r)=\left(\Omega_m(r)-
 \Omega_{m,\,{\rm out}}\right)/\Omega_{m,\,{\rm out}}$, where
 the subscript ``\,out\,'' indicates the value (far) outside the void of
 corresponding quantity. Following e.g.~\cite{Hoscheit:2018nfl,Cai:2020tpy},
 we assume that $\Omega_\Lambda(r)$ is constant. So, we have
 \be{eq12}
 \Omega_m(r)=\Omega_{m,\,{\rm out}}\left(1+\delta(r)\right)\,,
 \quad \Omega_\Lambda(r)=1-\Omega_{m,\,{\rm out}}\,,\quad
 \Omega_k(r)=1-\Omega_m(r)-\Omega_\Lambda(r)\,.
 \ee
 Note that slightly different boundary conditions on $\Omega_m(r)$ and
 $\Omega_\Lambda(r)$ have been chosen in~\cite{Kenworthy:2019qwq} (see
 its Appendix~A for details).~Nevertheless, in this work we use
 Eq.~(\ref{eq12}) following e.g.~\cite{Hoscheit:2018nfl,Cai:2020tpy},
 since they are extensively considered in the literature.~On the other
 hand, the boundary condition on the time since the Big Bang $t_B(r)$
 is equivalent to the condition on $H_0(r)$~\cite{GarciaBellido:2008nz,
 Garcia-Bellido:2008sdt,Yan:2014eca,Hoscheit:2018nfl,Kenworthy:2019qwq,
 Cai:2020tpy}.~In the constrained GBH (CGBH)
 model~\cite{GarciaBellido:2008nz,Garcia-Bellido:2008sdt} (see also
 e.g.~\cite{Yan:2014eca,Hoscheit:2018nfl,Kenworthy:2019qwq,
 Cai:2020tpy}), it is assumed that the Big Bang is spatially
 homogeneous, namely $t_B(r)=t_B={\rm const.}$, which can be
 set to $0$. Integrating Eq.~(\ref{eq7}), we find
 \bea
 \disp t-t_B=t=\frac{1}{H_0(r)}\int_0^R\frac{dR}{R}\left[\,
 \Omega_m(r)\,\frac{r^3}{R^3}+\Omega_k(r)\,\frac{r^2}{R^2}+
 \Omega_\Lambda(r)\,\right]^{-1/2}\,,\label{eq13}\\[1mm]
 \disp t_0-t_B=t_0=\frac{1}{H_0(r)}\int_0^r\frac{dR}{R}\left[
 \,\Omega_m(r)\,\frac{r^3}{R^3}+\Omega_k(r)\,\frac{r^2}{R^2}+
 \Omega_\Lambda(r)\,\right]^{-1/2}\,,\label{eq14}
 \eea
 in which we have used the gauge $R_0=r$ at $t_0$.~On the other
 hand, outside the void, it is a flat $\Lambda$CDM background
 FRW universe, which easily leads to
 \be{eq15}
 t_0-t_B=t_0=\int_0^1\frac{da}{H_{0,\,{\rm out}}\left[\,
 \Omega_{m,\,{\rm out}}\,a^{-1}+\left(1-\Omega_{m,\,{\rm out}}
 \right)a^2\,\right]^{\,1/2}}\,,
 \ee
 where $a$ plays the role of scale factor outside
 the void.~Following~\cite{Cai:2020tpy}, we require that
 Eqs.~(\ref{eq14}) and (\ref{eq15}) are equal, then obtain
 \be{eq16}
 H_0(r)=\left(t_0-t_B\right)^{-1}\int_0^1\frac{dx}{x}\left[
 \,\Omega_m(r)\,x^{-3}+\Omega_k(r)\,x^{-2}+\Omega_\Lambda(r)
 \,\right]^{-1/2}\,,
 \ee
 where $x=R/r$ and $t_0-t_B$ is given by Eq.~(\ref{eq15}).~Now we can
 numerically solve the differential equations
 $(\ref{eq7})-(\ref{eq9})$ with $H_0(r)$ given by Eq.~(\ref{eq16}) and
 $k(r)=-r^2 H_0^2(r)\,\Omega_k(r)/c^2$ from Eq.~(\ref{eq4}).~Then, using
 Eq.~(\ref{eq10}), the luminosity distance $d_L(z)$ is ready to
 be confronted with the observational data.


 \begin{table}[tb]
 \renewcommand{\arraystretch}{1.7}
 \begin{center}
 \vspace{-3mm}  
 \begin{tabular}{c} \hline\hline
 $\left|\,\Delta {\rm AIC}\,\right|$\\ \hline
 \renewcommand{\arraystretch}{1.7}
 \quad\ Level of empirical support for the model with the smaller AIC\quad\ \ \\ \hline
 \renewcommand{\arraystretch}{1.0}
 \begin{tabular}{ccc}\\[-3mm]
 $0-2$\hspace{10mm} &$4-7$\hspace{10mm} &$>10$\\
 Weak\hspace{10mm} &Mild\hspace{10mm} &Strong\\[1.7mm]
 \end{tabular}
 \\ \hline\hline
 $\left|\,\Delta {\rm BIC}\,\right|$\\ \hline
 \renewcommand{\arraystretch}{1.7}
 Evidence against the model with the larger BIC\\ \hline
 \renewcommand{\arraystretch}{1.0}
 \begin{tabular}{cccc}\\[-3mm]
 $0-2$\hspace{10mm} &$2-6$\hspace{10mm} &$6-10$\hspace{10mm} &$>10$\\
 Weak\hspace{10mm} &Positive\hspace{10mm} &Strong\hspace{10mm} &Very strong\\[1.7mm]
 \end{tabular}
 \\ \hline\hline
 $\left|\,\ln {\cal B}\,\right|$\\ \hline
 \renewcommand{\arraystretch}{1.7}
 Evidence against the model with the smaller $\cal Z$\\ \hline
 \renewcommand{\arraystretch}{1.0}
 \begin{tabular}{cccc}\\[-3mm]
 $0-1$\hspace{10mm} & $1-2.5$\hspace{10mm} & $2.5-5$\hspace{10mm} & $>5$\\
 Inconclusive\hspace{10mm} &Weak\hspace{10mm} & Moderate\hspace{10mm} & Strong\\[1.7mm]
 \end{tabular}
 \\ \hline\hline
 \end{tabular}
 \end{center}
 \vspace{-1mm}  
 \caption{\label{tab1} The empirical strength of
 $\Delta$AIC, $\Delta$BIC~\cite{Perivolaropoulos:2022khd} (see
 also e.g.~\cite{Perivolaropoulos:2023iqj,Liu:2024vlt}), and
 $\ln {\cal B}$~\cite{Kilbinger:2009by} (see also
 e.g.~\cite{Heavens:2017hkr}).}
 \end{table}



\subsection{The fiducial models and the information criteria}\label{sec2b}

The local $\Lambda$LTB void models should be compared with some fiducial
 models.~In the present work, the first fiducial model is a
 flat $\Lambda$CDM model in the FRW universe, which is given by
 \be{eq17}
 E(z)\equiv H(z)/H_0=\sqrt{\Omega_m\,(1+z)^3+(1-\Omega_m)}\,,\hspace{10mm}
 d_L(z)=(1+z)\,\frac{c}{H_0}\int_0^z\frac{d\tilde{z}}{E(\tilde{z})}\,,
 \ee
 and a constant absolute magnitude $M$ is needed to fit the
 SNIa data.~We label it as $\Lambda$CDM$_{\rm F}$ with three free model
 parameters $\Omega_m$, $H_0$ and $M$.~The second fiducial model is also
 a flat $\Lambda$CDM model, but its $\Omega_m=0.3153$ and $H_0=67.36\; {\rm
 km/s/Mpc}$ are fixed by Planck 2018 result~\cite{Planck:2018vyg}.~We label
 it as $\Lambda$CDM$_{\rm P18}$ with only one free model parameter $M$.

When we compare two models, $\Delta\chi^2$ is not enough, since
 they might have different numbers of free model parameters.~In the
 literature, a conventional criterion for model comparison is
 $\chi^2/dof$, where the degree of freedom $dof=N-\kappa$, while $N$ and
 $\kappa$ are the number of data points and the number of free model
 parameters, respectively. The most sophisticated criterion is
 the Bayesian evidence (see e.g.~\cite{Liddle:2007fy,Liddle:2009xe} and
 references therein).~But the computation of Bayesian evidence usually
 consumes a large amount of time and power.~As an alternative, some
 approximations of Bayesian evidence such as the information criteria
 AIC and BIC have been extensively used in the literature. The AIC is
 defined by~\cite{Akaike:1974}
 \be{eq18}
 {\rm AIC}=-2\ln {\cal L}_{\rm max}+2\kappa\,,
 \ee
 where ${\cal L}_{\rm max}$ is the maximum likelihood.~In the Gaussian
 cases, $\chi^2_{\rm min}=-2\ln {\cal L}_{\rm max}$.~The BIC is
 defined by~\cite{Schwarz:1978}
 \be{eq19}
 {\rm BIC}=-2\ln {\cal L}_{\rm max}+\kappa\ln N\,.
 \ee
 Comparing Eqs.~(\ref{eq18}) and (\ref{eq19}), it is easy to see that
 BIC is tougher than AIC for $\ln N>2$.~Notice that the Pantheon+ SNIa
 sample consists of 1701 data points and $\ln N\sim 7.44\gg
 2$.~Actually, this is the deep reason for the failures
 of~\cite{Perivolaropoulos:2023iqj,Liu:2024vlt} in terms of BIC, as
 mentioned above.~Usually, a smaller AIC or BIC means a better fitting
 for the given model, and we compare two models by calculating
 the differences in AIC and BIC. In this work, we will calculate $\Delta
 {\rm AIC}$ and $\Delta {\rm BIC}$ for the $\Lambda$LTB models relative
 to two fiducial $\Lambda$CDM models.~A negative (positive) value of
 $\Delta {\rm AIC}$ or $\Delta {\rm BIC}$ means a preference for the
 $\Lambda$LTB ($\Lambda$CDM) model.~Notice that the strength of evidence
 is indicated by the empirical ranges of $\left|\hspace{0.2mm}\Delta
 {\rm AIC}\hspace{0.3mm}\right|$ or $\left|\hspace{0.2mm}\Delta
 {\rm BIC}\hspace{0.3mm}\right|$ summarized
 in Table~\ref{tab1}~\cite{Perivolaropoulos:2022khd} (see also
 e.g.~\cite{Perivolaropoulos:2023iqj,Liu:2024vlt}).~Finally,
 we also consider the Bayesian evidence and the Bayes factor
 (see e.g.~\cite{Kass:1995loi,Kilbinger:2009by,Weinberg:2009rd,
 Trotta:2008qt,Mukherjee:2017oom}) for model comparison.~The Bayesian
 evidence is defined by
 \be{eqBy1}
 {\cal Z}=\int {\cal L}(\boldsymbol{\psi})\,P(\boldsymbol{\psi})\,d
 \boldsymbol{\psi}\,,
 \ee
 where $\cal L$ is the likelihood function, $P$ is the prior
 distribution, and $\boldsymbol{\psi}$ denotes the model parameters.~For
 model comparison, it is convenient to use the Bayes factor
 \be{eqBy2}
 {\cal B}_{12}={\cal Z}_1/{\cal Z}_2\,,\quad {\rm or~equivalently,}
 \quad\ln {\cal B}_{12}=\ln {\cal Z}_1 -\ln {\cal Z}_2\,,
 \ee
 where ${\cal Z}_1$ and ${\cal Z}_2$ are the Bayesian evidences for
 models $Q_1$ and $Q_2$, respectively.~If ${\cal B}_{12}$ is larger
 (smaller) than $1$, model $Q_1$ ($Q_2$) is preferred over the other
 model.~Here, we will also compute $\ln {\cal B}$ for the $\Lambda$LTB
 models relative to two fiducial $\Lambda$CDM models.~A
 positive (negative) value of $\ln {\cal B}$ means a preference for the
 $\Lambda$LTB ($\Lambda$CDM) model.~The strength of evidence is also
 indicated by the empirical ranges of $\left|\,\ln {\cal B}\,\right|$
 summarized in Table~\ref{tab1}~\cite{Kilbinger:2009by} (see
 also e.g.~\cite{Heavens:2017hkr}).~Note that one can compute
 the Bayesian evidence by using the nested sampling (such as PolyChord,
 dynesty, MultiNest, nessai), or alternatively MCEvidence with
 the MCMC chains~\cite{Heavens:2017afc,MCEvidence,MCEvimod}.


 \begin{center}
 \begin{figure}[tb]
 \centering
 \vspace{-4mm}  
 \includegraphics[width=0.58\textwidth]{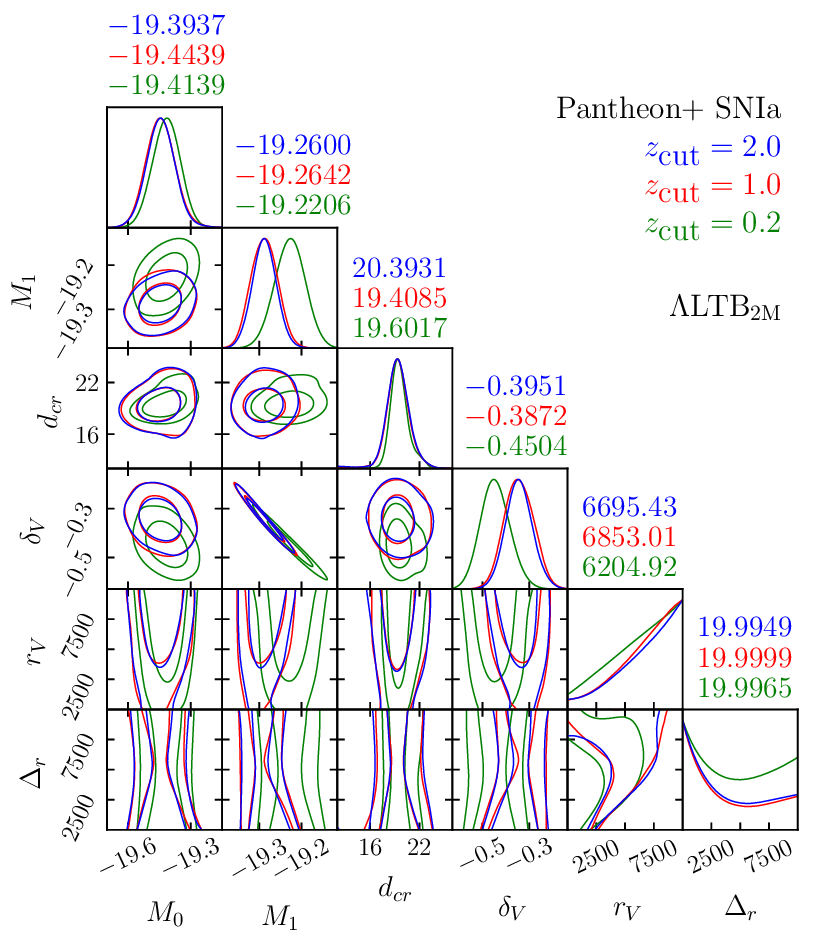}
 \vspace{-1mm}  
 \caption{\label{fig1} The $1\sigma$ and $2\sigma$ contours for all the
 free parameters of the $\Lambda$LTB model with a transition of the
 absolute magnitude $M$, i.e.~the $\Lambda$LTB$_{\rm 2M}$ model, from
 three $z<z_{\rm cut}$ subsets with $z_{\rm cut}=0.2$ (green),
 $1.0$ (red), $2.0$ (blue) of the Pantheon+ SNIa sample.~The
 marginalized probability distributions and the best-fit values are
 also given at the tops of all columns for the corresponding
 parameters.~The quantities of length e.g.~$d_{cr}$, $r_V$ and
 $\Delta_r$ are in units of $\rm Mpc$. See Sec.~\ref{sec3} for details.}
 \end{figure}
 \end{center}



 \begin{center}
 \begin{figure}[tb]
 \centering
 \vspace{-4mm}  
 \includegraphics[width=0.768\textwidth]{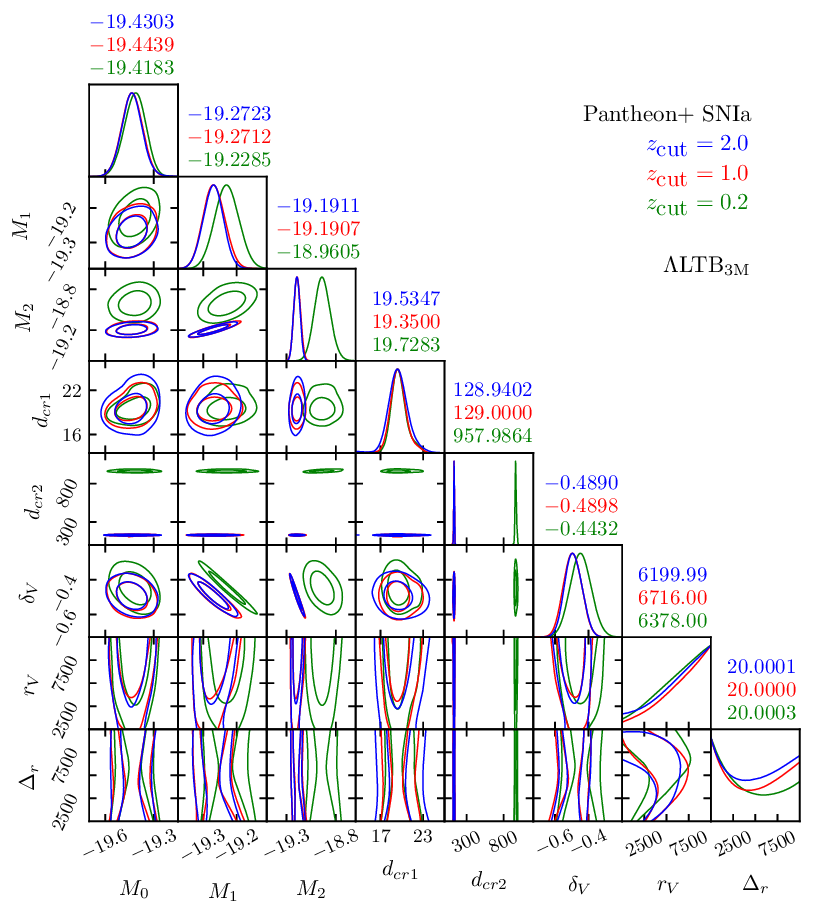}
 \vspace{-1mm}  
 \caption{\label{fig2} The same as in Fig.~\ref{fig1}, but for
 the $\Lambda$LTB$_{\rm 3M}$ model. See Sec.~\ref{sec3} for details.}
 \end{figure}
 \end{center}



 \begin{center}
 \begin{figure}[tb]
 \centering
 \vspace{-6mm}  
 \includegraphics[width=0.85\textwidth]{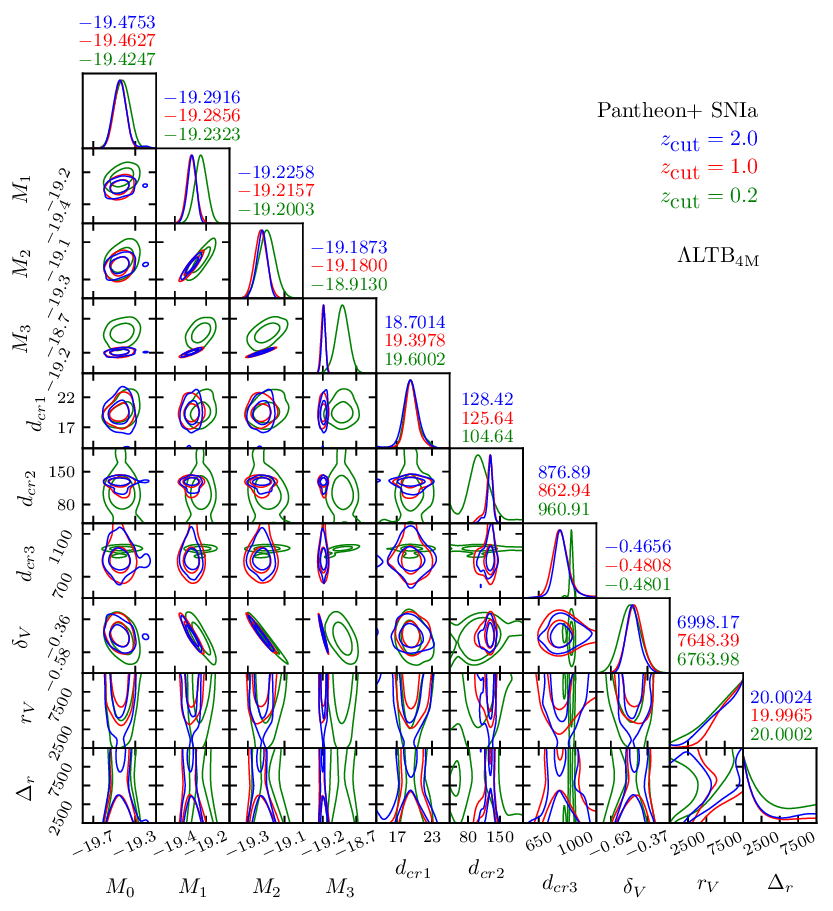}
 \vspace{-1mm}  
 \caption{\label{fig3} The same as in Fig.~\ref{fig1}, but for
 the $\Lambda$LTB$_{\rm 4M}$ model. See Sec.~\ref{sec3} for details.}
 \end{figure}
 \end{center}


\vspace{-25mm}  


\section{Testing the models with the Pantheon+ SNIa sample}\label{sec3}

In this work, we are interested in alleviating the Hubble tension with
 a local $\Lambda$LTB void.~To this end, the conditions
 $\Omega_{m,\,{\rm out}}=0.3153$ and $H_{0,\,{\rm out}}=67.36\;
 {\rm km/s/Mpc}$ for the $\Lambda$LTB models should be adopted to match
 the CMB observations of Planck 2018~\cite{Planck:2018vyg}, as
 in e.g.~\cite{Cai:2020tpy}.~So, the local universe can be described by
 a $\Lambda$LTB model inside the void (and outside the void nearby),
 while the background universe can be described by a flat
 $\Lambda$CDM$_{\rm P18}$ model far outside the void.~The cosmology
 smoothly transits between them through the $\tanh$ function in GBH
 profile given by Eq.~(\ref{eq11}).

The Pantheon+ SNIa sample~\cite{Brout:2022vxf,Scolnic:2021amr,
 PantheonPlusSH0ES} consists of 1701 light curves of 1550 SNIa
 at redshits $0.00122\leq z\leq 2.26137$ (in this work we use the Hubble
 diagram redshifts $z_{\rm HD}$ given by column 3 of the Pantheon+ data
 table with CMB and peculiar velocity corrections).~Only one of them
 is at redshift $z>2$.~Notice that 77 Cepheid calibrated host-galaxy
 distance moduli are also provided by SH0ES (many of them are
 duplicate) at redshifts $0.00122\leq z\leq 0.01682$ (only 10 of them
 are at $z>0.01$).

Following~\cite{Brout:2022vxf}, the cosmological parameters are
 constrained with Pantheon+ SNIa by minimizing
 \be{eq20}
 \chi^2_{\rm SN}=\Delta{\boldsymbol{\mu}}^{\,T}
 \cdot\boldsymbol{C}_{\rm stat+sys}^{\,-1}
 \cdot\Delta{\boldsymbol{\mu}}\,,
 \ee
 where $\boldsymbol{C}_{\rm stat+sys}$ is the covariance matrix
 including statistical and systematic uncertainties, $\Delta
 {\boldsymbol{\mu}}$ is the vector of 1701 SNIa distance
 modulus residuals computed as
 \be{eq21}
 \Delta\mu_i=
 \begin{cases}\\[-6.4mm]
 \;\mu_i-\mu_i^{\rm Ceph} & i\in {\rm Cepheid\ hosts},\\[1.7mm]
 \;\mu_i-\mu_{\rm model}(z_i) \quad & {\rm otherwise},
 \end{cases}
 \ee
 in which $\mu_i=m_{B,\,i}-M$ is the distance modulus of the $i$-th
 SNIa, $M$ is the absolute magnitude, $\mu_i^{\rm Ceph}$ is the
 Cepheid calibrated host-galaxy distance modulus provided by
 SH0ES, $m_{B,\,i}$ is the corrected/standardized apparent magnitude, and
 the model distance modulus $\mu_{\rm model}(z_i)$ is given by
 \be{eq22}
 \mu_{\rm model}(z_i)=5\log\left(d_L(z_i)/{\rm Mpc}\right)+25\,,
 \ee
 while $d_L(z)$ is the luminosity distance predicted by the model, and
 ``\,$\log$\,'' gives the logarithm to base $10$. We minimize
 $\chi^2_{\rm SN}$ by using the Markov Chain Monte Carlo (MCMC)
 Python package Cobaya~\cite{Torrado:2020dgo,Cobaya} with
 GetDist~\cite{Lewis:2019xzd,GetDist}, and then obtain
 the constraints on the model parameters.

Following~\cite{Cai:2020tpy}, we consider three subsets of the Pantheon+
 SNIa sample, namely three $z<z_{\rm cut}$ subsets with $z_{\rm cut}=
 0.2$, $1.0$, $2.0$, respectively.~Note that the case of $z<2.0$ is
 almost the full Pantheon+ sample, since there is only one SNIa
 at redshift $z>2$.~In the following, we test the models with these
 three SNIa subsets consisting of 948, 1676, 1700 data points, respectively.


 \begin{sidewaystable}[tbp] 
 \renewcommand{\arraystretch}{2.1}
 \begin{center}
 \vspace{-1mm}  
 \hspace{-1mm}  
 \begin{tabular}{lcccccccccc}\hline\hline
 Model & $M_0$ & $M_1$ & $M_2$ & $M_3$ & $d_{cr1}$ & $d_{cr2}$ & $d_{cr3}$ & $\delta_V$ & $r_V$ & $\Delta_r$\\[1mm] \hline
  & & & & & $z<0.2$ & & & & & \\[1mm] \hline
 $\Lambda$LTB$_{\rm 2M}$ & $-19.42^{+0.06}_{-0.06}$ & $-19.23^{+0.03}_{-0.03}$ & & & $19.57^{+0.59}_{-0.99}$ & & & $-0.44^{+0.05}_{-0.06}$ & $>5403.66$ & none\\[-2mm]
 $\Lambda$LTB$_{\rm 3M}$ & $-19.42^{+0.06}_{-0.05}$ & $-19.23^{+0.03}_{-0.03}$ & $-18.94^{+0.06}_{-0.07}$ & & $19.64^{+0.57}_{-1.04}$ & $962.92^{+7.69}_{-9.40}$ & & $-0.45^{+0.05}_{-0.05}$ & $>5431.18$ & none\\[-2mm]
 $\Lambda$LTB$_{\rm 4M}$ & $-19.42^{+0.06}_{-0.06}$ & $-19.23^{+0.03}_{-0.03}$ & $-19.20^{+0.04}_{-0.04}$ & $-18.91^{+0.09}_{-0.07}$ & $19.60^{+0.56}_{-1.05}$ & $103.63^{+21.91}_{-10.53}$ & $959.23^{+11.18}_{-5.55}$ & $-0.49^{+0.06}_{-0.06}$ & $>5763.84$ & none\\[1mm] \hline
  & & & & & $z<1.0$ & & & & & \\[1mm] \hline
 $\Lambda$LTB$_{\rm 2M}$ & $-19.44^{+0.05}_{-0.06}$ & $-19.29^{+0.03}_{-0.03}$ & & & $19.34^{+0.84}_{-0.89}$ & & & $-0.36^{+0.05}_{-0.06}$ & $>6573.41$ & none\\[-2mm]
 $\Lambda$LTB$_{\rm 3M}$ & $-19.43^{+0.06}_{-0.06}$ & $-19.27^{+0.03}_{-0.03}$ & $-19.19^{+0.03}_{-0.03}$ & & $19.61^{+0.58}_{-1.00}$ & $129.02^{+3.43}_{-3.58}$ & & $-0.49^{+0.05}_{-0.05}$ & $>5838.62$ & none\\[-2mm]
 $\Lambda$LTB$_{\rm 4M}$~ & ~$-19.45^{+0.06}_{-0.06}$~ & ~$-19.29^{+0.03}_{-0.03}$~ & ~$-19.23^{+0.03}_{-0.03}$~ & ~$-19.19^{+0.03}_{-0.03}$~ & ~$19.43^{+0.69}_{-0.91}$~ & ~$126.90^{+6.21}_{-2.26}$~ & ~$865.56^{+22.58}_{-70.62}$~ & ~$-0.46^{+0.05}_{-0.05}$~ & ~$>7091.30$~ & ~$<1699.04$~\\[1mm] \hline
  & & & & & $z<2.0$ & & & & & \\[1mm] \hline
 $\Lambda$LTB$_{\rm 2M}$ & $-19.44^{+0.06}_{-0.06}$ & $-19.29^{+0.02}_{-0.02}$ & & & $19.73^{+0.82}_{-0.81}$ & & & $-0.37^{+0.05}_{-0.05}$ & $>6401.97$ & none\\[-2mm]
 $\Lambda$LTB$_{\rm 3M}$ & $-19.44^{+0.05}_{-0.05}$ & $-19.27^{+0.03}_{-0.03}$ & $-19.19^{+0.03}_{-0.03}$ & & $19.45^{+0.71}_{-0.87}$ & $129.40^{+3.06}_{-3.84}$ & & $-0.49^{+0.04}_{-0.05}$ & $>5535.31$ & none\\[-2mm]
 $\Lambda$LTB$_{\rm 4M}$ & $-19.45^{+0.06}_{-0.06}$ & $-19.29^{+0.03}_{-0.03}$ & $-19.23^{+0.03}_{-0.03}$ & $-19.19^{+0.03}_{-0.03}$ & $19.13^{+1.01}_{-0.65}$ & $127.20^{+7.30}_{-2.06}$ & $879.04^{+16.57}_{-88.82}$ & $-0.46^{+0.05}_{-0.05}$ & $>6354.77$ & none\\[1mm] \hline\hline
 \end{tabular}
 \end{center}
 \vspace{-1mm}  
 \caption{\label{tabPost1} The means and $1\sigma$ uncertainties for
 all the free parameters of the $\Lambda$LTB models from the subsets
 $z<z_{\rm cut}$ with $z_{\rm cut}=0.2$, $1.0$, $2.0$ of the Pantheon+
 SNIa sample. Note that $d_{cr1}$ should be regarded as $d_{cr}$ in the
 $\Lambda$LTB$_{\rm 2M}$ model. The quantities of length
 e.g.~$d_{cr,\,i}$\,, $r_V$ and $\Delta_r$ are in units of $\rm Mpc$.
 In the last column, ``\,none\,'' indicates that the corresponding model
 parameter cannot be well constrained by the data. See
 Sec.~\ref{sec3} for details.}
 \end{sidewaystable}


For the following model comparison, we fit the fiducial
 $\Lambda$CDM$_{\rm F}$ model to three $z<z_{\rm cut}$ SNIa subsets with
 $z_{\rm cut}=0.2$, $1.0$, $2.0$, and find $\chi^2_{\rm SN,\,min,\,F}=
 901.516$, $1500.854$, $1522.907$, respectively.~For the
 fiducial $\Lambda$CDM$_{\rm P18}$ model, $\chi^2_{\rm SN,\,min,\,P18}=
 943.442$, $1546.216$, $1565.501$, respectively.

Actually, we have tried the $\Lambda$LTB model with a universal
 absolute magnitude $M$, and found that the Pantheon+ SNIa sample does
 not prefer it over the fiducial $\Lambda$CDM models in terms of AIC
 and BIC. On the other hand, hints for a transition of the
 absolute magnitude $M$ of SNIa have been found
 in~\cite{Perivolaropoulos:2023iqj,Liu:2024vlt}, as mentioned above.
 Motivated by this, here we consider the $\Lambda$LTB model
 with a transition of the absolute magnitude $M$.~Following
 \cite{Perivolaropoulos:2023iqj}, we assume a transition of $M$ at a
 critical distance $d_{cr}$ (associated with a critical distance modulus
 $\mu_{cr}=5\log\left(d_{cr}/{\rm Mpc}\right)+25$), namely
 \be{eq23}
 M=
 \begin{cases}
 \;M_0 \quad & {\rm if}\ \mu_{i,\,\rm S}<\mu_{cr}\,,\\[1.6mm]
 \;M_1 & {\rm if}\ \mu_{i,\,\rm S}\geq\mu_{cr}\,,
 \end{cases}
 \ee
 where $\mu_{i,\,\rm S}$ is given by column 11 of the Pantheon+ data
 table (note that it is roughly inferred by using the SH0ES
 absolute magnitude, and hence $\mu_{i,\,\rm S}$ cannot be used for
 cosmological constraints~\cite{Brout:2022vxf,Scolnic:2021amr,
 PantheonPlusSH0ES}).~We label this model as $\Lambda$LTB$_{\rm 2M}$.


 \begin{center}
 \begin{figure}[tb]
 \centering
 \vspace{-4mm}  
 \includegraphics[width=0.61\textwidth]{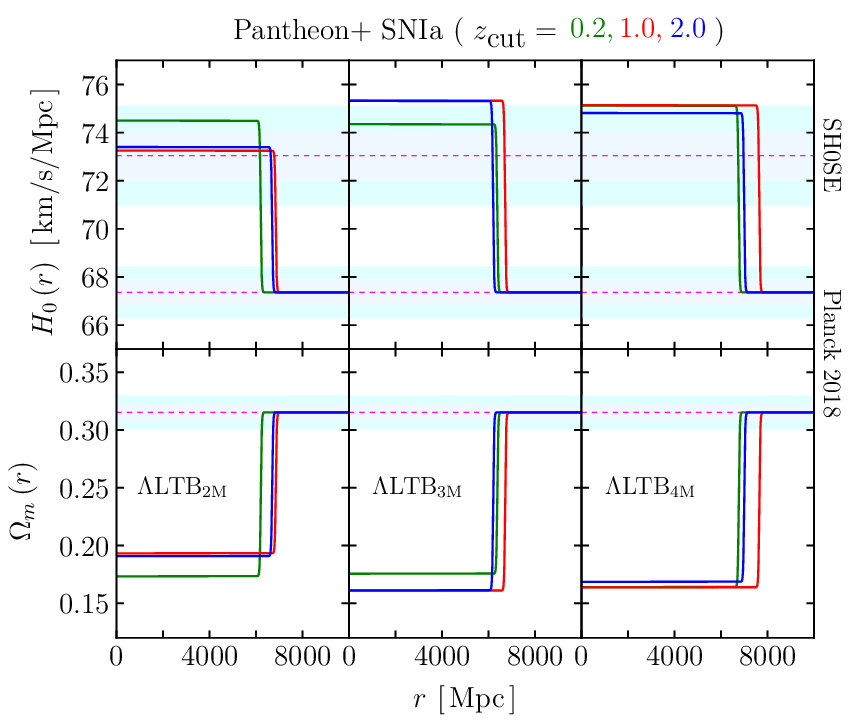}
 \vspace{-1mm}  
 \caption{\label{fig4} $H_0(r)$ and $\Omega_m(r)$ of the
 $\Lambda$LTB$_{\rm 2M}$ (left), $\Lambda$LTB$_{\rm 3M}$
 (middle), $\Lambda$LTB$_{\rm 4M}$ (right) void models as functions of
 the radial variable $r$ plotted with the best-fit parameters from three
 $z<z_{\rm cut}$ subsets with $z_{\rm cut}=0.2$ (green), $1.0$ (red),
 $2.0$ (blue) of the Pantheon+ SNIa sample.~The $1\sigma$, $2\sigma$
 uncertainties and median values of $H_0$ from SH0ES, $H_0$ and
 $\Omega_m$ from Planck 2018, are also shown by the shaded regions with
 dashed lines. See Sec.~\ref{sec3} for details.}
 \end{figure}
 \end{center}



 \begin{table}[tb]
 \renewcommand{\arraystretch}{1.7}
 \begin{center}
 \vspace{3mm}   
 \hspace{-1.63mm}  
 \begin{tabular}{lccccccrccr}\hline\hline
 Model & ~$\chi^2_{\rm SN,\,min}$~ & $\chi^2_{\rm SN,\,min}/dof$ & ~$\Delta {\rm BIC_F}$~
 & ~$\Delta {\rm AIC_F}$ & ~~~~$\Delta\chi^2_{\rm F}$~~~~ & $\ln {\cal B}_{\rm F}$ & ~$\Delta {\rm BIC_{P18}}$
 & ~$\Delta {\rm AIC_{P18}}$~ & $\Delta\chi^2_{\rm P18}$ & ~$\ln {\cal B}_{\rm P18}$\\ \hline
  & & & & & \hspace{-11.5mm}$z<0.2$ & & & & & \\ \hline
 $\Lambda$LTB$_{\rm 2M}$ & 887.96 &	0.9426 & 7.01 & $-7.55$ & $-13.55$ & 2.06 & $-21.21$ & $-45.48$ & $-55.48$ & 17.42\\[-2mm]
 $\Lambda$LTB$_{\rm 3M}$ & 854.01 & 0.9085 & $-13.23$ & $-37.50$ & $-47.50$ & 11.89 & $-41.45$ & $-75.43$ & $-89.43$ & 27.25\\[-2mm]
 $\Lambda$LTB$_{\rm 4M}$ & 847.55 &	0.9036 & $-5.99$ & $-39.97$ & $-53.97$ & 12.03 & $-34.20$ & $-77.89$ & $-95.89$ & 27.39\\ \hline
  & & & & & \hspace{-11.5mm}$z<1.0$ & & & & & \\ \hline
 $\Lambda$LTB$_{\rm 2M}$ & 1497.16 & 0.8965 & 18.58 & 2.31 & $-3.69$ & $-2.38$ & $-11.93$ & $-39.05$ & $-49.05$ & 14.10\\[-2mm]
 $\Lambda$LTB$_{\rm 3M}$ & 1468.56 & 0.8804 & 4.83 & $-22.29$ & $-32.29$ & 1.37 & $-25.68$ & $-63.65$ & $-77.65$ & 17.86\\[-2mm]
 $\Lambda$LTB$_{\rm 4M}$ & 1456.58 & 0.8743 & 7.70 & $-30.27$ & $-44.27$ & 3.44 & $-22.82$ & $-71.63$ & $-89.63$ & 19.93\\ \hline
  & & & & & \hspace{-11.5mm}$z<2.0$ & & & & & \\ \hline
 $\Lambda$LTB$_{\rm 2M}$ & 1515.38 & 0.8946 & 14.78 & $-1.53$ & $-7.53$ & $-0.16$ & $-12.93$ & $-40.12$ & $-50.12$ & 14.63\\[-2mm]
 $\Lambda$LTB$_{\rm 3M}$ & 1486.69 & 0.8787 & 0.98 & $-26.22$ & $-36.22$ & 3.58 & $-26.74$ & $-64.81$ & $-78.81$ & 18.37\\[-2mm]
 $\Lambda$LTB$_{\rm 4M}$\hspace{3mm} & 1474.91 & 0.8727 & 4.07 & $-34.00$ & $-48.00$ & 3.54 & $-23.65$ & $-72.60$ & $-90.60$ & 18.33\\ \hline\hline
 \end{tabular}
 \end{center}
 \vspace{-1mm}  
 \caption{\label{tab2} The model comparison of various
 $\Lambda$LTB void models relative to the fiducial $\Lambda$CDM$_{\rm
 F}$ and $\Lambda$CDM$_{\rm P18}$ models (labeled by the subscripts
 ``\,F\,'' and ``\,P18\,'' respectively), by using the subsets
 $z<z_{\rm cut}$ with $z_{\rm cut}=0.2$, $1.0$, $2.0$ of the
 Pantheon+ SNIa sample. See Sec.~\ref{sec3} for details.}
 \end{table}


\vspace{-10mm}  

It is natural to imagine that the absolute magnitude $M$ has
 many transitions, which are very reasonable generalizations.~In the
 $\Lambda$LTB void model, if $M$ have two transitions at the critical
 distances $d_{cr1}$ and~$d_{cr2}$ (associated with the
 critical distance moduli $\mu_{cr1,\,2}=5\log\left(d_{cr1,\,2}/{\rm
 Mpc}\right)+25$), namely
 \vspace{-0.1mm}  
 \be{eq24}
 M=
 \begin{cases}
 \;M_0 \quad & {\rm if}\ \mu_{i,\,\rm S}<\mu_{cr1}\,,\\[1.2mm]
 \;M_1 & {\rm if}\ \mu_{cr1}\leq\mu_{i,\,\rm S}<\mu_{cr2}\,,\\[1.2mm]
 \;M_2 & {\rm if}\ \mu_{i,\,\rm S}\geq\mu_{cr2}\,,
 \end{cases}
 \ee
 we label this model as $\Lambda$LTB$_{\rm 3M}$.~Similarly, if $M$ have
 three transitions at the critical distances $d_{cr1}$, $d_{cr2}$ and
 $d_{cr3}$ (associated with the critical distance moduli
 $\mu_{cr1,\,2,\,3}=5\log\left(d_{cr1,\,2,\,3}/{\rm Mpc}\right)+25$) in
 the $\Lambda$LTB void model, namely
 \be{eq25}
 M=
 \begin{cases}
 \;M_0 \quad & {\rm if}\ \mu_{i,\,\rm S}<\mu_{cr1}\,,\\[1.24mm]
 \;M_1 & {\rm if}\ \mu_{cr1}\leq\mu_{i,\,\rm S}<\mu_{cr2}\,,\\[1.24mm]
 \;M_2 & {\rm if}\ \mu_{cr2}\leq\mu_{i,\,\rm S}<\mu_{cr3}\,,\\[1.24mm]
 \;M_3 & {\rm if}\ \mu_{i,\,\rm S}\geq\mu_{cr3}\,,
 \end{cases}
 \ee
 we label it as $\Lambda$LTB$_{\rm 4M}$. More transitions of $M$ are not
 worthy, and hence we stop here.

We fit the $\Lambda$LTB$_{\rm 2M}$, $\Lambda$LTB$_{\rm 3M}$,
 $\Lambda$LTB$_{\rm 4M}$ void models to three $z<z_{\rm cut}$ subsets
 with $z_{\rm cut}=0.2$, $1.0$, $2.0$ of the Pantheon+ SNIa sample, and
 present the $1\sigma$ and $2\sigma$ constraints on their free model
 parameters in Figs.~$\ref{fig1}-\ref{fig3}$, respectively.~Notice that
 the best-fit values of their model parameters are also given in these
 figures.~Although the marginalized probability distributions and the
 $1\sigma$, $2\sigma$ contours of all the free model parameters have
 been plotted in Figs.~$\ref{fig1}-\ref{fig3}$, we also explicitly give
 their numerical means and $1\sigma$ intervals in Table~\ref{tabPost1}.~On
 the other hand, we present their best-fit $\chi^2_{\rm SN,\,min}$ and
 $\chi^2_{\rm SN,\,min}/dof$ in Table~\ref{tab2}. Then, we calculate
 their $\Delta\chi^2$, $\Delta {\rm AIC}$, $\Delta {\rm BIC}$,
 $\ln {\cal B}$ relative to the fiducial $\Lambda$CDM$_{\rm F}$ and
 $\Lambda$CDM$_{\rm P18}$ models (labeled by the subscripts ``\,F\,''
 and ``\,P18\,'' respectively), and also present them in
 Table~\ref{tab2}.~In Fig.~\ref{fig4}, we plot $H_0(r)$ and $\Omega_m(r)$ as
 functions of the comoving distance $r$ by using the best-fit
 model parameters given in Figs.~$\ref{fig1}-\ref{fig3}$.

From Fig.~\ref{fig4}, one can easily find the local Hubble constant
 $H_{0,\,{\rm in}}=H_0(r=0)$ at the center $r=0$ (where we live) for the
 $\Lambda$LTB void models, which will be compared with the one measured
 by SH0ES (as shown by the upper shaded regions with a dashed line in
 Fig.~\ref{fig4}).~Far outside the void, it is clear to see that all
 the $H_0(r)$ at large $r$ of three $\Lambda$LTB models match
 $H_{0,\,{\rm out}}=67.36\;{\rm km/s/Mpc}$ from the CMB observation of
 Planck 2018 (as shown by the lower shaded regions with a dashed line
 in Fig.~\ref{fig4}).~Actually, this is the very condition imposed at
 the beginning of Sec.~\ref{sec3}.~Obviously, in the $\Lambda$LTB$_{\rm
 2M}$ model, its local $H_{0,\,{\rm in}}=H_0(r=0)$ perfectly closes to
 the median value $H_0=73.04\;{\rm km/s/Mpc}$ of SH0ES for the cases of
 $z_{\rm cut}=1.0$, $2.0$, and falls into $2\sigma$ region of SH0ES
 in the case of $z_{\rm cut}=0.2$.~On the other hand, in the
 $\Lambda$LTB$_{\rm 3M}$ and $\Lambda$LTB$_{\rm 4M}$ models,
 their $H_{0,\,{\rm in}}=H_0(r=0)$ fall into $2\sigma$ region of SH0ES
 in all cases.~So, all the three $\Lambda$LTB void models can
 significantly alleviate the Hubble tension, as expected.

From Figs.~$\ref{fig1}-\ref{fig3}$ and Table~\ref{tabPost1}, we can see
 that the characteristic radius $r_V$ and transition width $\Delta_r$ of
 the voids cannot be well constrained by the Pantheon+ SNIa sample.~Note
 that in all the three $\Lambda$LTB void models the best-fit $\Delta_r
 \sim 20\;{\rm Mpc}\ll r_V$ is fairly short, and hence $\Omega_m(r)$ and
 $H_0(r)$ transit the edge of the local void around $r_V$ steeply, as
 shown in Fig.~\ref{fig4}.~On the other hand, the void depth (fractional
 deficit) $\delta_V$ can be well constrained in all the three
 $\Lambda$LTB void models, and they can be $\delta_V\lesssim -38\%$,
 even $\lesssim -48\%$, as shown in Figs.~$\ref{fig1}-\ref{fig3}$ and
 Table~\ref{tabPost1}.~This also can be seen from Fig.~\ref{fig4} that
 the local $\Omega_{m,\,{\rm in}}=\Omega_m(r=0)<0.2$ is about
 $\left(1+\delta_V\right)\Omega_{m,\,{\rm out}}=
 \left(1-38\%\right)\times 0.3153\sim 0.2$ or even smaller.~As noted
 in~\cite{Cai:2020tpy}, $\delta_V\lesssim -30\%$ is required to
 alleviate the Hubble tension, but they failed by using the Pantheon
 SNIa sample.~With the help of transitions of the absolute magnitude
 $M$, we can satisfyingly make $\delta_V\lesssim -38\%$ or even smaller
 in the present work by using the Pantheon+ SNIa sample.~This is the
 key to alleviate the Hubble tension with a local void.

On the other hand, from Figs.~$\ref{fig1}-\ref{fig3}$ we find
 a transition of the absolute magnitude $M$ at $\sim 20\;{\rm Mpc}$
 in all models.~Thus, we independently confirm this point found
 in~\cite{Perivolaropoulos:2023iqj} previously.~In the present work, we
 further find two transitions of the absolute magnitude $M$ at
 $\sim 129\;{\rm Mpc}$ and $\sim 860-960\;{\rm Mpc}$.~Meanwhile, we
 find in all the three $\Lambda$LTB void models that the segmented $M_i$
 becomes higher as the distance increases (namely $M_3>M_2>M_1>M_0$
 at the critical distances $d_{cr3}>d_{cr2}>d_{cr1}$), as shown
 in Figs.~$\ref{fig1}-\ref{fig3}$.~So, if the absolute magnitude $M=M(d)$ or
 $M(z)$, it might be a (monotone) increasing function.

From Table~\ref{tabPost1}, we can see that the constraints on the free
 model parameters are fairly tight (except $r_V$ and $\Delta_r$).~We
 stress that the transitions of the absolute magnitude $M$ are statistically
 significant (far beyond $1\sigma$ uncertainties).~They are not apparent
 changes due to large errors.~On the other hand, despite the void parameters
 $r_V$ and $\Delta_r$ cannot be well constrained, they do not affect the
 main results since the key void parameter $\delta_V$ can be tightly
 constrained in fact.

From Table~\ref{tab2}, it is easy to see that all the three $\Lambda$LTB
 models are strongly preferred over the fiducial $\Lambda$CDM$_{\rm
 P18}$ model with ``\,very strong\,'' $\Delta\chi^2_{\rm P18}$, $\Delta
 {\rm AIC_{P18}}$, $\Delta{\rm BIC_{P18}}$ and $\ln {\cal B}_{\rm P18}$
 evidences.~On the other hand, the $\Lambda$LTB$_{\rm 3M}$ and
 $\Lambda$LTB$_{\rm 4M}$ models are preferred over the fiducial
 $\Lambda$CDM$_{\rm F}$ model with ``\,very strong\,'' $\Delta
 {\rm AIC_F}$ and $\Delta\chi^2_{\rm F}$ evidences in all cases, while
 the $\Lambda$LTB$_{\rm 2M}$ model is mildly preferred over the fiducial
 $\Lambda$CDM$_{\rm F}$ model.~However, they fail in many cases with
 positive $\Delta {\rm BIC_F}$ relative to the fiducial
 $\Lambda$CDM$_{\rm F}$ model, as shown by the 4th column of
 Table~\ref{tab2}.~So, in terms of BIC, the $\Lambda$LTB void models are
 still not preferred over the fiducial $\Lambda$CDM$_{\rm F}$ model, by
 using the Pantheon+ SNIa sample alone.~Even in terms of $\ln
 {\cal B}_{\rm F}$, the $\Lambda$LTB void models are only preferred over the
 fiducial $\Lambda$CDM$_{\rm F}$ model with ``\,moderate\,'' evidences
 in many cases, and fail with weakly negative $\ln {\cal B}_{\rm F}$ in
 two cases.~Thus, we still cannot claim a full triumph so far.


 \begin{center}
 \begin{figure}[tb]
 \centering
 \vspace{-3mm}  
 \includegraphics[width=0.58\textwidth]{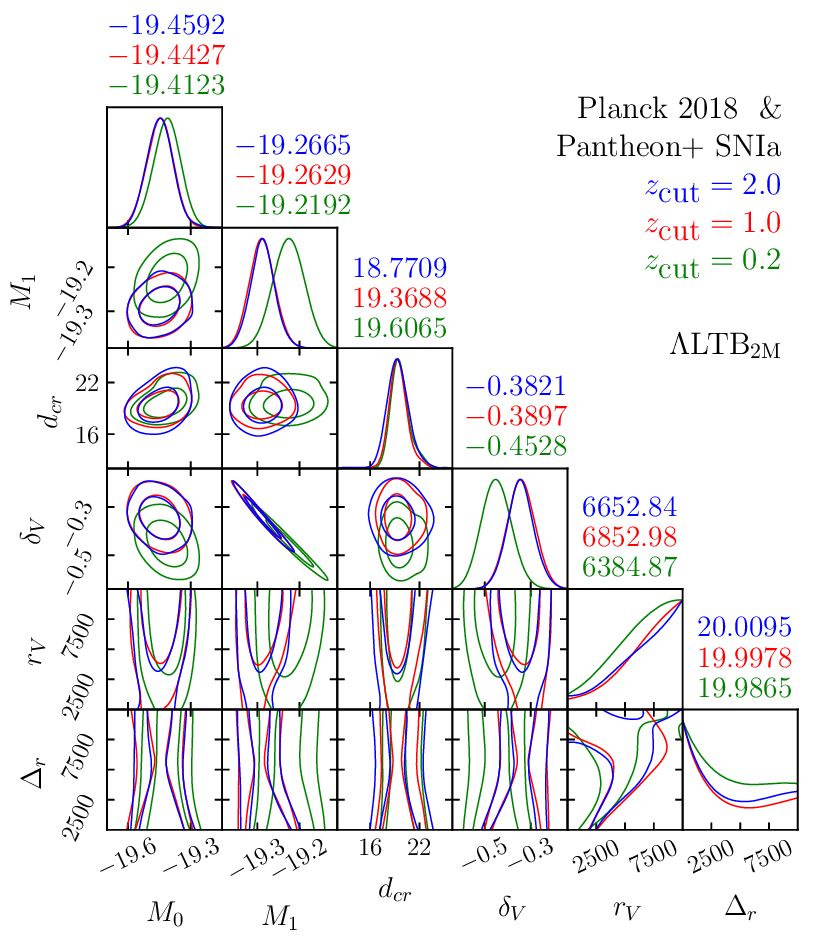}
 \vspace{-1mm}  
 \caption{\label{fig5} The same as in Fig.~\ref{fig1}, but for
 the $\Lambda$LTB$_{\rm 2M}$ model and the data of SNIa+CMB.
 See Sec.~\ref{sec4} for details.}
 \end{figure}
 \end{center}



 \begin{center}
 \begin{figure}[tb]
 \centering
 \vspace{-5mm}  
 \includegraphics[width=0.768\textwidth]{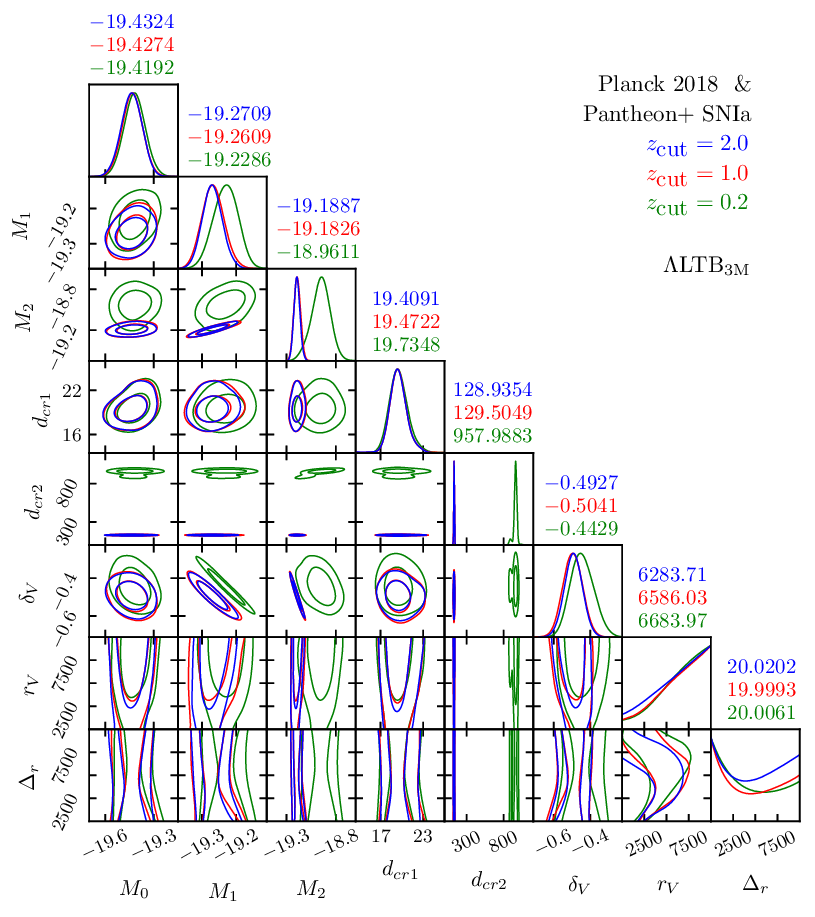}
 \vspace{-1mm}  
 \caption{\label{fig6} The same as in Fig.~\ref{fig1}, but for
 the $\Lambda$LTB$_{\rm 3M}$ model and the data of SNIa+CMB.
 See Sec.~\ref{sec4} for details.}
 \end{figure}
 \end{center}



 \begin{center}
 \begin{figure}[tb]
 \centering
 \vspace{-6mm} 
 \includegraphics[width=0.85\textwidth]{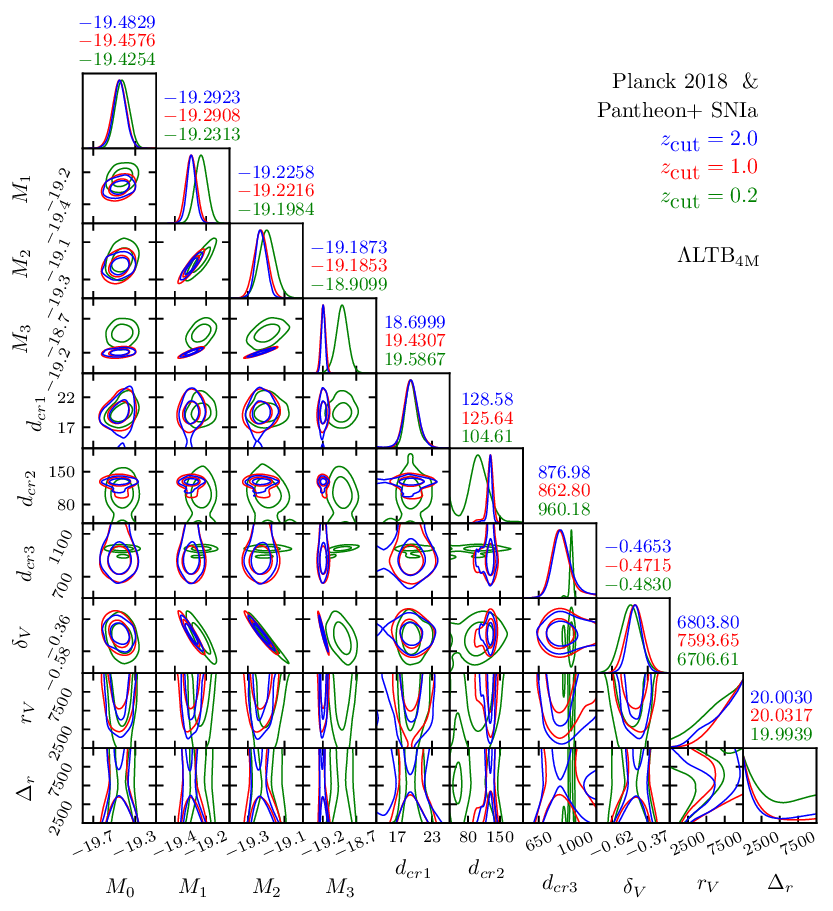}
 \vspace{-1mm}  
 \caption{\label{fig7} The same as in Fig.~\ref{fig1}, but for
 the $\Lambda$LTB$_{\rm 4M}$ model and the data of SNIa+CMB.
 See Sec.~\ref{sec4} for details.}
 \end{figure}
 \end{center}


\vspace{-24mm} 


\section{Testing the models with SNIa and CMB}\label{sec4}

It is worth noting that the conditions $\Omega_{m,\,{\rm out}}=0.3153$
 and $H_{0,\,{\rm out}}=67.36\;{\rm km/s/Mpc}$ for the $\Lambda$LTB
 models have been imposed with the strong intention of reconciling the
 discrepancy between the local observation of SNIa and the
 high-$z$ observation of CMB, as mentioned at the beginning of
 Sec.~\ref{sec3}.~If one only uses the SNIa data, it is not necessary to
 impose these conditions on $\Omega_{m,\,{\rm out}}$ and $H_{0,\,{\rm
 out}}$, and they should be regarded as free model parameters.~It is
 unfair to impose these conditions without using the CMB data.~In other
 words, when we impose these conditions on $\Omega_{m,\,{\rm out}}$ and
 $H_{0,\,{\rm out}}$, the CMB data must be also taken into account,
 jointly with the SNIa data.

It consumes a large amount of time and power to use the full CMB data.~As an
 alternative, the distance priors derived from the full CMB data have been
 extensively used the literature, which contain the main information of
 CMB.~Here, we consider the distance priors in~\cite{Chen:2018dbv} derived
 from the CMB observation of Planck 2018.~In this case, the $\chi^2$ for CMB
 is given by~\cite{Chen:2018dbv}
 \be{eq26}
 \chi^2_{\rm CMB}=\Delta\boldsymbol{d}^{\,T}
 \cdot\boldsymbol{C}_{d}^{\,-1}\cdot\Delta\boldsymbol{d}\,,
 \ee
 where $\Delta\boldsymbol{d}$ is the vector of distance prior
 residuals, and $\boldsymbol{C}_{d}^{\,-1}$ is the inverse covariance
 matrix, namely
 \be{eq27}
 \Delta\boldsymbol{d}=\left(\begin{array}{c}
                              {\cal R}_{\rm model}(z_\ast)-1.750235\\
                              \ell_{A,\,{\rm model}}(z_\ast)-301.4707
                            \end{array}\right)\,,\hspace{10mm}
 \boldsymbol{C}_{d}^{\,-1}=\left(\begin{array}{lr}
                                   94392.3971 & \ -1360.4913\\
                                   -1360.4913 & 161.4349
                            \end{array}\right)\,,
 \ee
 in which $z_\ast=1089.92$ from the Planck 2018
 result~\cite{Planck:2018vyg}.~The shift parameter $\cal R$ is given by
 \be{eq28}
 {\cal R}_{\rm model}(z_\ast)=
 \left(H_0/c\right)\sqrt{\Omega_m}\,\left(1+z_\ast\right)d_A(z_\ast)\,,
 \ee
 and the acoustic scale $\ell_A$ is given by
 \be{eq29}
 \ell_{A,\,{\rm model}}(z_\ast)=\left(1+z_\ast\right)\pi d_A(z_\ast)
 /r_s(z_\ast)\,,
 \ee
 where $d_A(z)=(1+z)^{-2}d_L(z)$ is the angular diameter distance, which
 can be obtained from the luminosity distance $d_L(z)$.~The comoving
 sound horizon $r_s(z)$ is given by~\cite{Chen:2018dbv}
 \be{eq30}
 r_s(z)=\frac{c}{H_0}\int_0^{1/(1+z)}\frac{da}{a^2 E(a)
 \,\sqrt{3\left(1+\eta a\right)}}\,,\quad
 \eta=\frac{3\Omega_b h^2}{4\Omega_\gamma h^2}=
 31500\left(T_{\rm CMB}/2.7\,{\rm K}\right)^{-4}\Omega_b h^2\,,
 \ee
 where $a=(1+z)^{-1}$ is the scale factor,
 $T_{\rm CMB}=2.7255\,{\rm K}$~\cite{Fixsen:2009ug}, and
 $\Omega_b h^2=0.02237$ from the Planck 2018
 result~\cite{Planck:2018vyg}.~Noting that the integration in
 Eq.~(\ref{eq30}) is computed at very high redshift $z\geq
 z_\ast=1089.92$, the radiation cannot be ignored.~In this case, the
 dimensionless Hubble parameter is given by~\cite{Chen:2018dbv}
 \bea
 &E(a)=\left[\,\Omega_r\,a^{-4}+\,\Omega_m\,a^{-3}+\left(1-\Omega_m
 -\Omega_r\right)\,\right]^{1/2}\,,\label{eq31}\\[1mm]
 &\Omega_r=\Omega_m/(1+z_{\rm eq})\,,\quad
 z_{\rm eq}=2.5\times 10^4 \left(T_{\rm CMB}/2.7\,{\rm K}\right)^{-4}
 \Omega_m h^2\,,\label{eq32}
 \eea
 where $h$ is $H_0$ in units of $100\;{\rm km/s/Mpc}$.~We minimize the
 total $\chi^2$, namely
 \be{eq33}
 \chi^2_{\rm tot} = \chi^2_{\rm SN} + \chi^2_{\rm CMB}\,,
 \ee
 to constrain the model parameters by using the Pantheon+ SNIa
 sample and the CMB data of Planck 2018 jointly.


 \begin{sidewaystable}[tbp] 
 \renewcommand{\arraystretch}{2.1}
 \begin{center}
 \vspace{-1mm}  
 \hspace{-1mm}  
 \begin{tabular}{lcccccccccc}\hline\hline
 Model & $M_0$ & $M_1$ & $M_2$ & $M_3$ & $d_{cr1}$ & $d_{cr2}$ & $d_{cr3}$ & $\delta_V$ & $r_V$ & $\Delta_r$\\[1mm] \hline
  & & & & & $z<0.2$ & & & & & \\[1mm] \hline
 $\Lambda$LTB$_{\rm 2M}$ & $-19.42^{+0.06}_{-0.05}$ & $-19.23^{+0.03}_{-0.03}$ & & & $19.66^{+0.50}_{-1.01}$ & & & $-0.45^{+0.05}_{-0.06}$ & $>5439.50$ & $<5466.16$\\[-2mm]
 $\Lambda$LTB$_{\rm 3M}$ & $-19.42^{+0.06}_{-0.06}$ & $-19.23^{+0.03}_{-0.03}$ & $-18.95^{+0.08}_{-0.07}$ & & $20.11^{+0.09}_{-1.54}$ & $958.81^{+10.10}_{-4.73}$ & & $-0.44^{+0.05}_{-0.06}$ & $>5697.92$ & none\\[-2mm]
 $\Lambda$LTB$_{\rm 4M}$ & $-19.42^{+0.06}_{-0.06}$ & $-19.23^{+0.03}_{-0.03}$ & $-19.19^{+0.04}_{-0.04}$ & $-18.91^{+0.08}_{-0.07}$ & $19.66^{+0.56}_{-1.10}$ & $104.04^{+26.42}_{-12.22}$ & $959.43^{+10.41}_{-5.71}$ & $-0.49^{+0.06}_{-0.06}$ & $>5941.53$ & none\\[1mm] \hline
  & & & & & $z<1.0$ & & & & & \\[1mm] \hline
 $\Lambda$LTB$_{\rm 2M}$ & $-19.45^{+0.06}_{-0.06}$ & $-19.29^{+0.03}_{-0.03}$ & & & $19.40^{+0.73}_{-0.89}$ & & & $-0.37^{+0.05}_{-0.05}$ & $>6183.68$ & none\\[-2mm]
 $\Lambda$LTB$_{\rm 3M}$ & $-19.43^{+0.05}_{-0.06}$ & $-19.27^{+0.03}_{-0.03}$ & $-19.19^{+0.03}_{-0.03}$ & & $19.53^{+0.64}_{-0.92}$ & $129.45^{+2.99}_{-3.99}$ & & $-0.50^{+0.04}_{-0.05}$ & $>5599.65$ & none\\[-2mm]
 $\Lambda$LTB$_{\rm 4M}$~ & ~$-19.45^{+0.06}_{-0.06}$~ & ~$-19.29^{+0.03}_{-0.03}$~ & ~$-19.23^{+0.03}_{-0.03}$~ & ~$-19.19^{+0.03}_{-0.03}$~ & ~$19.34^{+0.82}_{-0.84}$~ & ~$127.34^{+5.59}_{-2.38}$~ & ~$869.07^{+19.68}_{-73.61}$~ & ~$-0.46^{+0.05}_{-0.06}$~ & ~$>7154.88$~ & ~$<1303.50$~\\[1mm] \hline
  & & & & & $z<2.0$ & & & & & \\[1mm] \hline
 $\Lambda$LTB$_{\rm 2M}$ & $-19.44^{+0.05}_{-0.06}$ & $-19.29^{+0.02}_{-0.02}$ & & & $19.24^{+0.73}_{-0.96}$ & & & $-0.37^{+0.05}_{-0.05}$ & $>5885.65$ & none\\[-2mm]
 $\Lambda$LTB$_{\rm 3M}$ & $-19.44^{+0.06}_{-0.06}$ & $-19.27^{+0.02}_{-0.03}$ & $-19.19^{+0.02}_{-0.03}$ & & $19.42^{+0.71}_{-0.83}$ & $129.41^{+2.94}_{-3.77}$ & & $-0.49^{+0.04}_{-0.04}$ & $>5158.87$ & none\\[-2mm]
 $\Lambda$LTB$_{\rm 4M}$ & $-19.45^{+0.06}_{-0.06}$ & $-19.29^{+0.03}_{-0.03}$ & $-19.23^{+0.03}_{-0.03}$ & $-19.19^{+0.03}_{-0.03}$ & $19.24^{+0.98}_{-0.83}$ & $127.64^{+4.25}_{-2.41}$ & $885.56^{+12.70}_{-99.33}$ & $-0.46^{+0.05}_{-0.05}$ & $>6148.70$ & none\\[1mm] \hline\hline
 \end{tabular}
 \end{center}
 \vspace{-1mm}  
 \caption{\label{tabPost2} The same as in Table~\ref{tabPost1}, but for
 the data of SNIa+CMB. See Sec.~\ref{sec4} for details.}
 \end{sidewaystable}


Note that the $\Lambda$LTB void model converges with the flat
 $\Lambda$CDM$_{\rm P18}$ model far outside the void.~Thus, when we
 compute the distance priors $\cal R$ and $\ell_A$ for the $\Lambda$LTB
 void model, $\Omega_m=\Omega_{m,\,{\rm out}}=0.3153$ and $H_0=
 H_{0,\,{\rm out}}=67.36\;{\rm km/s/Mpc}$ should be used, since the CMB
 scale is certainly far outside the void. On the other hand, in the
 fiducial $\Lambda$CDM$_{\rm F}$ model, both $\Omega_m$ and $H_0$ are
 free model parameters.


 \begin{center}
 \begin{figure}[tb]
 \centering
 \vspace{-4mm}  
 \includegraphics[width=0.61\textwidth]{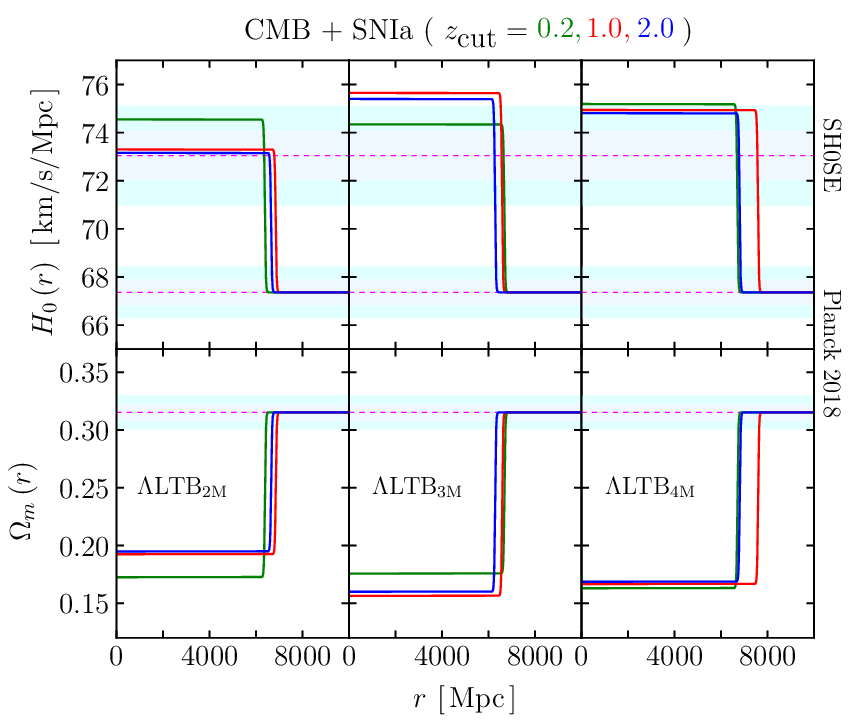}
 \vspace{-1mm}  
 \caption{\label{fig8} The same as in Fig.~\ref{fig4}, but for the data
 of SNIa+CMB. See Sec.~\ref{sec4} for details.}
 \end{figure}
 \end{center}



 \begin{table}[tb]
 \renewcommand{\arraystretch}{1.7}
 \begin{center}
 \vspace{3mm}   
 \hspace{-1.63mm}  
 \begin{tabular}{lccccccrccr}\hline\hline
 Model & ~$\chi^2_{\rm tot,\,min}$~ & $\chi^2_{\rm tot,\,min}/dof$ & ~$\Delta {\rm BIC_F}$~
 & ~$\Delta {\rm AIC_F}$ & ~~~~$\Delta\chi^2_{\rm F}$~~~~ & $\ln {\cal B}_{\rm F}$ & ~$\Delta {\rm BIC_{P18}}$
 & ~$\Delta {\rm AIC_{P18}}$~ & $\Delta\chi^2_{\rm P18}$ & ~$\ln {\cal B}_{\rm P18}$\\ \hline
  & & & & & \hspace{-11.5mm}$z<0.2$ & & & & & \\ \hline
 $\Lambda$LTB$_{\rm 2M}$ & 893.06 & 0.9460 & $-25.14$ & $-39.71$ & $-45.71$ & 23.96 & $-16.48$ & $-40.77$ & $-50.77$ & 15.68\\[-2mm]
 $\Lambda$LTB$_{\rm 3M}$ & 859.01 & 0.9119 & $-45.47$ & $-69.75$ & $-79.75$ & 34.16 & $-36.82$ & $-70.81$ & $-84.81$ & 25.87\\[-2mm]
 $\Lambda$LTB$_{\rm 4M}$ & 852.55 & 0.9070 & $-38.23$ & $-72.22$ & $-86.22$ & 33.65 & $-29.58$ & $-73.28$ & $-91.28$ & 25.36\\ \hline
  & & & & & \hspace{-11.5mm}$z<1.0$ & & & & & \\ \hline
 $\Lambda$LTB$_{\rm 2M}$ & 1500.37 & 0.8973 & $-20.77$ & $-37.05$ & $-43.05$ & 21.50 & $-9.11$ & $-36.24$ & $-46.24$ & 12.00\\[-2mm]
 $\Lambda$LTB$_{\rm 3M}$ & 1471.44 & 0.8811 & $-34.85$ & $-61.97$ & $-71.97$ & 25.12 & $-23.18$ & $-61.16$ & $-75.16$ & 15.62\\[-2mm]
 $\Lambda$LTB$_{\rm 4M}$ & 1460.76 & 0.8758 & $-30.68$ & $-68.65$ & $-82.65$ & 27.03 & $-19.01$ & $-67.84$ & $-85.84$ & 17.53\\ \hline
  & & & & & \hspace{-11.5mm}$z<2.0$ & & & & & \\ \hline
 $\Lambda$LTB$_{\rm 2M}$ & 1518.86 & 0.8956 & $-21.03$ & $-37.35$ & $-43.35$ & 22.06 & $-9.83$ & $-37.02$ & $-47.02$ & 12.83\\[-2mm]
 $\Lambda$LTB$_{\rm 3M}$ & 1490.80 & 0.8800 & $-34.20$ & $-61.40$ & $-71.40$ & 25.59 & $-23.01$ & $-61.08$ & $-75.08$ & 16.35\\[-2mm]
 $\Lambda$LTB$_{\rm 4M}$\hspace{3mm} & 1479.91 & 0.8747 & $-30.22$ & $-68.30$ & $-82.30$ & 25.70 & $-19.02$ & $-67.98$ & $-85.98$ & 16.47\\ \hline\hline
 \end{tabular}
 \end{center}
 \vspace{-1mm}  
 \caption{\label{tab3} The same as in Table~\ref{tab2}, but for the data
 of SNIa+CMB. See Sec.~\ref{sec4} for details.}
 \end{table}


\vspace{-11mm} 

For the model comparison, we fit the fiducial $\Lambda$CDM$_{\rm F}$
 model to the CMB data of Planck 2018 and three $z<z_{\rm cut}$ SNIa
 subsets with $z_{\rm cut}=0.2$, $1.0$, $2.0$, then find $\chi^2_{\rm
 tot,\,min,\,F}=938.769$, $1543.413$, $1562.208$, respectively.~For the
 fiducial $\Lambda$CDM$_{\rm P18}$ model, $\chi^2_{\rm tot,\,min,\,P18}=
 943.827$, $1546.602$, $1565.886$, respectively.

We fit the $\Lambda$LTB$_{\rm 2M}$, $\Lambda$LTB$_{\rm 3M}$,
 $\Lambda$LTB$_{\rm 4M}$ models to the same CMB+SNIa datasets
 with $z_{\rm cut}=0.2$, $1.0$, $2.0$, and present the $1\sigma$ and
 $2\sigma$ constraints on their free model parameters in
 Figs.~$\ref{fig5}-\ref{fig7}$, respectively. Notice that the best-fit
 values of their model parameters are also given in these
 figures.~Although the marginalized probability distributions and the
 $1\sigma$, $2\sigma$ contours of all the free model parameters have
 been plotted in Figs.~$\ref{fig5}-\ref{fig7}$, we also explicitly give
 their numerical means and $1\sigma$ intervals in
 Table~\ref{tabPost2}.~On the other hand, we present their best-fit
 $\chi^2_{\rm tot,\,min}$ and $\chi^2_{\rm tot,\,min}/dof$ in
 Table~\ref{tab3}.~Then, we calculate their $\Delta\chi^2$, $\Delta
 {\rm AIC}$, $\Delta {\rm BIC}$, $\ln {\cal B}$ relative to the fiducial
 $\Lambda$CDM$_{\rm F}$ and $\Lambda$CDM$_{\rm P18}$ models (labeled by
 the subscripts ``\,F\,'' and ``\,P18\,'' respectively), and
 also present them in Table~\ref{tab3}.~In Fig.~\ref{fig8}, we plot $H_0(r)$
 and $\Omega_m(r)$ as functions of the comoving distance $r$ by using
 the best-fit model parameters given in Figs.~$\ref{fig5}-\ref{fig7}$.

From Fig.~\ref{fig8}, it is easy to see that the Hubble tension can
 be satisfyingly saved in the $\Lambda$LTB$_{\rm 2M}$ void model, and
 can be significantly alleviated in the $\Lambda$LTB$_{\rm 3M}$ and
 $\Lambda$LTB$_{\rm 4M}$ void models within $2\sigma$ confidence level
 (CL.), mainly due to the fairly deep void depth
 $\delta_V\lesssim -38\%$, even $\lesssim -50\%$, as shown in
 Figs.~$\ref{fig5}-\ref{fig7}$ (and supported by $\Omega_{m,\,{\rm in}}=
 \Omega_m(r=0)<0.2$ as shown in Fig.~\ref{fig8}).

Again, we confirm the transition of the absolute magnitude
 $M$ at $\sim 20\;{\rm Mpc}$ found in~\cite{Perivolaropoulos:2023iqj}
 previously, and find two transitions of the absolute magnitude $M$
 at $\sim 129\;{\rm Mpc}$ and $\sim 860-960\;{\rm Mpc}$, as shown in
 Figs.~$\ref{fig5}-\ref{fig7}$. The segmented $M_i$ increases with the
 distance (namely $M_3>M_2>M_1>M_0$ at the critical distances $d_{cr3}>
 d_{cr2}>d_{cr1}$), as shown in Figs.~$\ref{fig5}-\ref{fig7}$.

From Table~\ref{tabPost2}, we can see that the constraints on the free
 model parameters are fairly tight (except $r_V$ and $\Delta_r$).~We
 stress that the transitions of the absolute magnitude $M$ are statistically
 significant (far beyond $1\sigma$ uncertainties).~They are not apparent
 changes due to large errors.~On the other hand, despite the void parameters
 $r_V$ and $\Delta_r$ cannot be well constrained, they do not affect the
 main results since the key void parameter $\delta_V$ can be tightly
 constrained in fact.

From Table~\ref{tab3}, it is easy to see that all the three $\Lambda$LTB
 models are super-strongly preferred over both fiducial
 $\Lambda$CDM$_{\rm F}$ and $\Lambda$CDM$_{\rm P18}$ models in all terms
 of $\Delta{\rm BIC}$, $\Delta{\rm AIC}$, $\Delta\chi^2$ and $\ln {\cal
 B}$ with ``\,very strong\,'' evidences, by using the data of SNIa+CMB.
 Clearly, these landslide preferences are overwhelming.

Let us look Table~\ref{tab3} closely.~In terms of $\Delta{\rm BIC}$
 ($\Delta{\rm AIC}$), the $\Lambda$LTB$_{\rm 3M}$ ($\Lambda$LTB$_{\rm
 4M}$) model is preferred over the other two $\Lambda$LTB models by the
 data of SNIa+CMB, respectively. In terms of $\ln {\cal B}$,
 the $\Lambda$LTB$_{\rm 3M}$ and $\Lambda$LTB$_{\rm 4M}$ models are
 comparable.~In all terms of $\Delta{\rm BIC}$, $\Delta{\rm AIC}$ and
 $\ln {\cal B}$, the $\Lambda$LTB$_{\rm 3M}$ and $\Lambda$LTB$_{\rm
 4M}$ models are strongly preferred over the $\Lambda$LTB$_{\rm 2M}$
 model.~In total, $\Lambda {\rm LTB_{3M}}\sim\Lambda {\rm LTB_{4M}}\gg
 \Lambda {\rm LTB_{2M}}$.~The $\Lambda$LTB void model with two
 transitions of the absolute magnitude $M$ is the best.~This hints at
 least one more transition of $M$ at $\sim 129\;{\rm Mpc}$ in addition
 to the one at $\sim 20\;{\rm Mpc}$ found in~\cite{Perivolaropoulos:2023iqj}
 previously.~Although the $\Lambda$LTB$_{\rm 2M}$ model is not the best,
 we stress that it is still overwhelmingly preferred over both fiducial
 $\Lambda$CDM$_{\rm F}$ and $\Lambda$CDM$_{\rm P18}$ models by
 the data of SNIa+CMB.


\section{Conclusion and discussions}\label{sec5}

Nowadays, one of the well-known serious challenges in cosmology is the
 Hubble tension, namely the discrepancy between the Hubble constants
 from the local observation of Type Ia supernova (SNIa) and the high-$z$
 observation of cosmic microwave background (CMB). Here, we are
 interested in alleviating the Hubble tension with a local void.~The key
 idea is to violate the cosmological principle by assuming that we live
 in a locally underdense void centered nearby our location.~In this
 underdense region, one will feel a locally faster expansion
 rate compared to the cosmic average. In the literature, it was found
 that a local void cannot satisfyingly alleviate the Hubble tension,
 since it is not preferred over the $\Lambda$CDM model by the
 observations such as the Pantheon SNIa sample, especially in terms of
 the information criteria AIC and BIC. In this work, we try to alleviate
 the Hubble tension with a local void and  transitions of the absolute
 magnitude $M$, by using the Pantheon+ SNIa sample alone or jointly with
 the CMB data of Planck 2018.~We find that the Hubble tension can be
 satisfyingly alleviated, while the $\Lambda$LTB models are strongly
 preferred by the observations.

As mentioned in Secs.~\ref{sec1} and \ref{sec3}, the key
 difference between the Pantheon+ SNIa sample and the other SNIa samples
 (e.g.~DESY5, Union3, Pantheon, JLA, Union2.1, SNLS3) is the 77 Cepheid
 calibrated host-galaxy distance moduli provided by SH0ES, which can be
 used to constrain the absolute magnitude $M$ alone (n.b.~the first row
 of Eq.~(\ref{eq21}), regardless of $H_0$ appeared in $d_L$ of
 the theoretical model), and hence the degeneracy between $H_0$ and $M$
 are broken, so that they can be separately constrained.~Thus, it is
 possible to simultaneously study the Hubble tension and transitions of
 the absolute magnitude $M$ by using the Pantheon+ SNIa sample.~Without
 these Cepheid calibrated host-galaxy distance moduli provided by SH0ES,
 however, $H_0$ and $M$ are heavily degenerated in the cases of the
 DESY5 and Union3 SNIa samples (through a combination ${\cal M}=M+5\log
 \left(c/H_0/{\rm Mpc}\right)+25$, which can be marginalized as
 a nuisance parameter).~In order to study the Hubble tension, usually in
 the literature one has to fix $M$ to be a constant value given by SH0ES
 or impose a Gaussian prior on $M$.~But in this way, we cannot further
 consider transitions of the absolute magnitude $M$.~Therefore, it is
 fairly difficult to use alternative SNIa samples (such as DESY5 and
 Union3) to independently replicate the same/similar works as in the
 present paper.

In this work, we have considered a phenomenological scenario combining
 a local void with multiple transitions of the absolute magnitude $M$,
 which significantly increases model complexity, and the
 risk is over-fitting the data without a compelling physical mechanism
 driving these transitions (we thank the referee for pointing out this
 issue).~First, the Occam's Razor is quantified by the Bayesian
 evidence, AIC and BIC. The added complexity is worthy if it
 can significantly improve the Bayesian evidence, AIC and BIC.~As shown
 in Tables~\ref{tab2} and \ref{tab3} (especially the latter), the reward
 of increasing model complexity is very great, while the evidences of
 AIC, BIC and $\ln {\cal B}$ are overwhelmingly strong.~Second, as shown
 in Tables~\ref{tabPost1} and \ref{tabPost2}, the transitions of the
 absolute magnitude $M$ are statistically significant (far beyond $1\sigma$
 uncertainties).~They are not apparent changes due to large errors and
 noises.~Finally, there might be some physical mechanisms for transitions of
 the absolute magnitude $M$.~In fact, the transition of $M$ at $\sim
 20\;{\rm Mpc}$ previously found in~\cite{Perivolaropoulos:2023iqj}
 is partly due to the volumetric redshift scatter
 bias~\cite{Perivolaropoulos:2023iqj,Kenworthy:2022jdh} (see
 also e.g.~\cite{Brout:2022vxf}).~However, it cannot completely explain
 the transition of $M$ at $\sim 20\;{\rm Mpc}$, as mentioned
 in~\cite{Perivolaropoulos:2023iqj}.~Thus, possible physical causes are
 needed for all the transitions of $M$ at $\sim 20\;{\rm Mpc}$, $\sim
 129\;{\rm Mpc}$ and $\sim 860-960\;{\rm Mpc}$.~For instance, they might
 be caused by the transitions of effective gravitational
 ``\,constant\,'' $G_{\rm eff}$~\cite{Marra:2021fvf,Alestas:2021nmi,
 Sapone:2020wwz,Caldwell:2005xb,Khosravi:2017hfi,Alestas:2021luu}, a
 broken symmetron screening mechanism~\cite{Perivolaropoulos:2022txg},
 or a sharp feature in scalar-tensor theory
 potential~\cite{Perivolaropoulos:2023iqj}. In particular, as is well known,
 a varying $G_{\rm eff}$ is common in many modified gravity theories. It can
 lead to a varying/modified Chandrasekhar limit of compact
 stars (e.g.~\cite{Wei:2021xek,Liu:2024vlt}). Since SNIa explosions are
 closely related to the Chandrasekhar limit of white dwarfs, it
 is reasonable to consider possible transitions of the SNIa absolute
 magnitude $M$.~Actually, there are other possible physical mechanisms
 to change the Chandrasekhar limit and then the SNIa absolute magnitude
 $M$, for example, a varying mass of electron, neutron or proton, a
 varying speed of light $c$, or a varying fine-structure
 ``\,constant\,'' $\alpha$, due to the possible coupling between the
 electromagnetic field and dark energy (e.g.~an evolving scalar
 field $\varphi$)~\cite{Wei:2009xg,Wei:2017mzf}.~In the theoretical
 market, more physical mechanisms are waiting us to this end. Let us
 keep an open mind.

A local void considered here violates the cosmological principle, which
 assumes that the universe is homogeneous and isotropic on
 cosmic scales.~However, the cosmological principle has not yet been
 well proven on cosmic scales
 $\gtrsim 1\;{\rm Gpc}$~\cite{Caldwell:2007yu}.~On the other hand, the
 local universe is obviously inhomogeneous and anisotropic on smaller
 scales.~In the literature, the cosmic homogeneity has been tested by
 using e.g.~SNIa, CMB, BAO, galaxy surveys, integrated
 Sachs-Wolfe effect, kinetic Sunyaev Zel'dovich effect, and large-scale
 structure (see e.g.~\cite{Yan:2014eca,Deng:2018yhb,Deng:2018jrp,Yu:2019cku}
 and references therein).~However, the debate on the inhomogeneous universe
 has not been settled by now~\cite{Celerier:1999hp,Celerier:2007jc,
 Barrett:1999fd,Tomita:2000jj,Tomita:2001gh,Iguchi:2001sq,Zhang:2012qr,
 Wang:2011kj,Enqvist:2006cg,Enqvist:2007vb,
 GarciaBellido:2008nz,Garcia-Bellido:2008sdt,Celerier:2011zh,
 Celerier:2012xr,Alnes:2005rw,Celerier:2009sv,
 Vanderveld:2006rb,Zibin:2008vj,Zibin:2011ptm,Yan:2014eca,Yu:2019cku,
 Alnes:2006pf,Sundell:2015cza,ChirinosIsidro:2016vah,Keenan:2013mfa,
 Hoscheit:2018nfl,Shanks:2018rka,Kenworthy:2019qwq,Lukovic:2019ryg,
 Kazantzidis:2020tko,Cai:2020tpy,Haslbauer:2020xaa,Mazurenko:2023sex,
 Mazurenko:2024gwj,Banik:2025dlo,
 Haridasu:2024ask,Aluri:2022hzs,Krishnan:2021jmh,McConville:2023xav,
 Boubel:2024cmh,Moffat:2016dbd,Moffat:1991qj,Moffat:1994qy,Moffat:2025whe,
 Castello:2021uad,Camarena:2022iae}.~The $\Lambda$LTB void
 firstly proposed in 1933 is still theoretically viable.~On the other
 hand, one might be concerned with the depth of void.~As noted
 in~\cite{Cai:2020tpy}, $\delta_V\lesssim -30\%$ is required to
 alleviate the Hubble tension.~Actually, we find $\delta_V
 \lesssim -38\%$ even $\lesssim -50\%$ in this work.~So,
 $\Omega_{m,\,{\rm in}}=\left(1+\delta_V\right)\Omega_{m,\,{\rm out}}
 \sim\left(1-38\%\right)\times 0.3153\sim 0.2$. In fact, a
 loose constraint on $\Omega_m$ from galaxy cluster survey is given by
 $\Omega_m=0.20\pm 0.04\pm 0.09$ with $1\sigma$ internal and systematic
 errors~\cite{Carlberg:1995aq}.~Thus, a deep local void is
 still consistent with large-scale structure observations.

In this work, we consider three subsets of the Pantheon+ SNIa sample
 with $z<z_{\rm cut}=0.2$, $1.0$, $2.0$ following~\cite{Cai:2020tpy}.~We
 also consider step-like transitions $M_i$ at critical
 distances $d_{cr,\,j}$, and the number of $M_i$ is chosen to be $i=0$,
 $1$, $2$, $3$, while the number of critical distances $d_{cr,\,j}$ is
 correspondingly chosen to be $j=1$, $2$, $3$.~The values of $M_i$ and
 $d_{cr,\,j}$ are determined by fitting to the data.~It is interesting
 to let $z_{\rm cut}$ float freely, and the numbers of $M_i$
 and $d_{cr,\,j}$ are sampled over in the MCMC analysis (we thank the
 referee for pointing out this issue).~But this is fairly difficult.~Notice
 that $z_{\rm cut}$ is not a model parameter. It is just used to determine
 which SNIa will be taken into account, while $z_{\rm cut}$ does not
 directly appear in the likelihood or $\chi^2$.~If $z_{\rm cut}$ floats
 freely, the number of SNIa at $z<z_{\rm cut}$ used to constrain the
 model parameters also floats accordingly.~On the other hand, if the numbers
 of $M_i$ and $d_{cr,\,j}$ are also not predefined, some new techniques
 are needed, since the number of model parameters is not known in this
 case.~A possible way out is to consider a series of crowded $M_i$,
 $d_{cr,\,j}$ (e.g.~$j=i+1=1$, $2$, $\dots$, $10$), and
 $z_{\rm cut}$ (e.g.~$0.1$, $0.2$, $0.3$, $\dots$, $2.0$).~Of course,
 it is better to instead use some new techniques to this end. But they
 are beyond our scope, and in order to save the length of the present
 paper, we leave them to the future works.


\section*{ACKNOWLEDGEMENTS}

We thank the anonymous referee for quite useful comments and
 suggestions, which helped us to improve this work.~We are grateful
 to Profs.~Puxun~Wu, Shao-Jiang~Wang, and Drs.~Wang-Wei~Yu, Yang~Liu, as
 well as Han-Yue~Guo, Lin-Yu~Li, Shu-Yan~Long, Hui-Qiang~Liu,
 Yu-Xuan~Li, Shuo-Yu~Zhang for kind help and useful discussions. This
 work was supported in part by NSFC under Grants No.~12375042
 and No.~11975046.

\renewcommand{\baselinestretch}{1.1}



\begin{thebibliography}{999}

\bibitem{DiValentino:2025sru}
E.~Di Valentino \textit{et al.},
Phys. Dark Univ. \textbf{49}, 101965 (2025)
[arXiv:2504.01669].

\bibitem{Abdalla:2022yfr}
E.~Abdalla \textit{et al.},
JHEAp \textbf{34}, 49 (2022)
[arXiv:2203.06142].

\bibitem{Perivolaropoulos:2021jda}
L.~Perivolaropoulos and F.~Skara,
New Astron. Rev. \textbf{95}, 101659 (2022)
[arXiv:2105.05208].

\bibitem{Verde:2023lmm}
L.~Verde, N.~Sch\"oneberg and H.~Gil-Mar\'\i{}n,
Ann. Rev. Astron. Astrophys. \textbf{62}, 287 (2024)
[arXiv:2311.13305].

\bibitem{DiValentino:2022fjm}
E.~Di Valentino,
Universe \textbf{8}, no.8, 399 (2022).

\bibitem{Efstathiou:2024dvn}
G.~Efstathiou,
Phil. Trans. Roy. Soc. Lond. A \textbf{383}, no.2290, 20240022 (2025)
[arXiv:2406.12106].

\bibitem{DiValentino:2021izs}
E.~Di Valentino \textit{et al.},
Class. Quant. Grav. \textbf{38}, no.15, 153001 (2021)
[arXiv:2103.01183].

\bibitem{Rong-Gen:2023dcz}
R.~G.~Cai, L.~Li and S.~J.~Wang,
Acta Phys. Sin. \textbf{72}, no.23, 239801 (2023).

\bibitem{Hu:2023jqc}
J.~P.~Hu and F.~Y.~Wang,
Universe \textbf{9}, no.2, 94 (2023)
[arXiv:2302.05709].

\bibitem{Chang:2022tzj}
C.~L.~Chang \textit{et al.},
arXiv:2203.07638 [astro-ph.CO].

\bibitem{Planck:2018vyg}
N.~Aghanim \textit{et al.},
Astron. Astrophys. \textbf{641}, A6 (2020)
[arXiv:1807.06209].

\bibitem{Riess:2021jrx}
A.~G.~Riess \textit{et al.},
Astrophys. J. Lett. \textbf{934}, no.1, L7 (2022)
[arXiv:2112.04510].

\bibitem{Riess:2024vfa}
A.~G.~Riess \textit{et al.},
Astrophys. J. \textbf{977}, no.1, 120 (2024)
[arXiv:2408.11770].

\bibitem{Lemaitre:1933gd}
  G.~Lema\^{\i}tre,
  Annales de la Soci\'et\'e Scientifique de Bruxelles
  A {\bf 53}, 51 (1933);\\
  see Gen.\ Rel.\ Grav.\  {\bf 29}, 641 (1997) for English translation.

\bibitem{Tolman:1934za}
  R.~C.~Tolman,
  Proc.\ Nat.\ Acad.\ Sci.\  {\bf 20}, 169 (1934);\\
  see Gen.\ Rel.\ Grav.\  {\bf 29}, 935 (1997) for English translation.

\bibitem{Bondi:1947fta}
  H.~Bondi, Mon.\ Not.\ Roy.\ Astron.\ Soc.\ {\bf 107}, 410 (1947).

\bibitem{Zehavi:1998gz}
I.~Zehavi, A.~G.~Riess, R.~P.~Kirshner and A.~Dekel,
Astrophys. J. \textbf{503}, 483 (1998)
[astro-ph/9802252].

\bibitem{Celerier:1999hp}
M.~N.~Celerier,
Astron. Astrophys. \textbf{353}, 63 (2000)
[astro-ph/9907206].

\bibitem{Celerier:2007jc}
M.~N.~Celerier,
New Advances in Physics \textbf{1}, 29 (2007)
[astro-ph/0702416].

\bibitem{Barrett:1999fd}
  R.~K.~Barrett and C.~A.~Clarkson,
  Class.\ Quant.\ Grav.\  {\bf 17}, 5047 (2000)
  [astro-ph/9911235].

\bibitem{Tomita:2000jj}
  K.~Tomita,
  Mon.\ Not.\ Roy.\ Astron.\ Soc.\  {\bf 326}, 287 (2001)
  [astro-ph/0011484].

\bibitem{Tomita:2001gh}
  K.~Tomita,
  Prog.\ Theor.\ Phys.\  {\bf 106}, 929 (2001)
  [astro-ph/0104141].

\bibitem{Iguchi:2001sq}
 H.~Iguchi, T.~Nakamura and K.~Nakao,
 Prog.\ Theor.\ Phys.\  {\bf 108}, 809 (2002)
 [astro-ph/0112419].

\bibitem{Zhang:2012qr}
  T.~J.~Zhang, H.~Wang and C.~Ma,
  Phys. Rev. D \textbf{91}, 063506 (2015)
  [arXiv:1210.1775].

\bibitem{Wang:2011kj}
  H.~Wang and T.~J.~Zhang,
  Astrophys.\ J.\  {\bf 748}, 111 (2012)
  [arXiv:1111.2400].

\bibitem{Enqvist:2006cg}
  K.~Enqvist and T.~Mattsson,
  JCAP {\bf 0702}, 019 (2007)
  [astro-ph/0609120].

\bibitem{Enqvist:2007vb}
  K.~Enqvist,
  Gen.\ Rel.\ Grav.\  {\bf 40}, 451 (2008)
  [arXiv:0709.2044].

\bibitem{GarciaBellido:2008nz}
  J.~Garcia-Bellido and T.~Haugboelle,
  JCAP {\bf 0804}, 003 (2008)
  [arXiv:0802.1523].

\bibitem{Garcia-Bellido:2008sdt}
J.~Garcia-Bellido and T.~Haugboelle,
JCAP \textbf{0809}, 016 (2008)
[arXiv:0807.1326].

\bibitem{Celerier:2011zh}
  M.~N.~Celerier,
  Astron.\ Astrophys.\  {\bf 543}, A71 (2012)
  [arXiv:1108.1373].

\bibitem{Celerier:2012xr}
  M.~N.~Celerier,
  J.\ Phys.\ Conf.\ Ser.\  {\bf 484}, 012005 (2014)
  [arXiv:1203.2814].

\bibitem{Alnes:2005rw}
  H.~Alnes, M.~Amarzguioui and O.~Gron,
  Phys.\ Rev.\ D {\bf 73}, 083519 (2006)
  [astro-ph/0512006].

\bibitem{Celerier:2009sv}
  M.~N.~Celerier {\it et al.},
  Astron.\ Astrophys.\  {\bf 518}, A21 (2010)
  [arXiv:0906.0905].

\bibitem{Vanderveld:2006rb}
  R.~A.~Vanderveld {\it et al.},
  Phys.\ Rev.\ D {\bf 74}, 023506 (2006)
  [astro-ph/0602476].

\bibitem{Zibin:2008vj}
  J.~P.~Zibin,
  Phys.\ Rev.\ D {\bf 78}, 043504 (2008)
  [arXiv:0804.1787].

\bibitem{Zibin:2011ptm}
J.~P.~Zibin and A.~Moss,
Class. Quant. Grav. \textbf{28}, 164005 (2011)
[arXiv:1105.0909].

\bibitem{Yan:2014eca}
X.~P.~Yan, D.~Z.~Liu and H.~Wei,
Phys. Lett. B \textbf{742}, 149 (2015)
[arXiv:1411.6218].

\bibitem{Yu:2019cku}
Z.~X.~Yu, S.~L.~Li and H.~Wei,
Nucl. Phys. B \textbf{960}, 115179 (2020)
[arXiv:1907.12517].

\bibitem{Alnes:2006pf}
H.~Alnes and M.~Amarzguioui,
Phys. Rev. D \textbf{74}, 103520 (2006)
[astro-ph/0607334].

\bibitem{Sundell:2015cza}
P.~Sundell, E.~M\"ortsell and I.~Vilja,
JCAP \textbf{1508}, 037 (2015)
[arXiv:1503.08045].

\bibitem{ChirinosIsidro:2016vah}
E.~G.~Chirinos Isidro {\it et al.},
JCAP \textbf{1605}, 003 (2016)
[arXiv:1602.08583].

\bibitem{Keenan:2013mfa}
R.~C.~Keenan, A.~J.~Barger and L.~L.~Cowie,
Astrophys. J. \textbf{775}, 62 (2013)
[arXiv:1304.2884].

\bibitem{Hoscheit:2018nfl}
B.~L.~Hoscheit and A.~J.~Barger,
Astrophys. J. \textbf{854}, no.1, 46 (2018)
[arXiv:1801.01890].

\bibitem{Shanks:2018rka}
T.~Shanks, L.~Hogarth and N.~Metcalfe,
Mon. Not. Roy. Astron. Soc. \textbf{484}, L64 (2019)
[arXiv:1810.02595].

\bibitem{Kenworthy:2019qwq}
W.~D.~Kenworthy, D.~Scolnic and A.~Riess,
Astrophys. J. \textbf{875}, no.2, 145 (2019)
[arXiv:1901.08681].

\bibitem{Lukovic:2019ryg}
V.~V.~Lukovi\'c {\it et al.},
Mon. Not. Roy. Astron. Soc. \textbf{491}, no.2, 2075 (2020)
[arXiv:1907.11219].

\bibitem{Kazantzidis:2020tko}
L.~Kazantzidis and L.~Perivolaropoulos,
Phys. Rev. D \textbf{102}, no.2, 023520 (2020)
[arXiv:2004.02155].

\bibitem{Cai:2020tpy}
R.~G.~Cai {\it et al.},
Phys. Rev. D \textbf{103}, no.12, 123539 (2021)
[arXiv:2012.08292].

\bibitem{Pan-STARRS1:2017jku}
D.~M.~Scolnic \textit{et al.},
Astrophys. J. \textbf{859}, no.2, 101 (2018)
[arXiv:1710.00845].

\bibitem{Brout:2022vxf}
D.~Brout \textit{et al.},
Astrophys. J. \textbf{938}, no.2, 110 (2022)
[arXiv:2202.04077].

\bibitem{Scolnic:2021amr}
D.~Scolnic \textit{et al.},
Astrophys. J. \textbf{938}, no.2, 113 (2022)
[arXiv:2112.03863].

\bibitem{PantheonPlusSH0ES}
https:$/\!/$PantheonPlusSH0ES.github.io \\ https:$/\!/$github.com/PantheonPlusSH0ES/DataRelease

\bibitem{Sorrenti:2022zat}
F.~Sorrenti, R.~Durrer and M.~Kunz,
JCAP \textbf{2311}, 054 (2023)
[arXiv:2212.10328].

\bibitem{Sorrenti:2024ztg}
F.~Sorrenti, R.~Durrer and M.~Kunz,
JCAP \textbf{2504}, 013 (2025)
[arXiv:2403.17741].

\bibitem{Sorrenti:2024ugq}
F.~Sorrenti, R.~Durrer and M.~Kunz,
JCAP \textbf{2412}, 003 (2024)
[arXiv:2407.07002].

\bibitem{Lopes:2024vfz}
M.~Lopes, A.~Bernui, C.~Franco and F.~Avila,
Astrophys. J. \textbf{967}, no.1, 47 (2024)
[arXiv:2405.11077].

\bibitem{Watkins:2023rll}
R.~Watkins \textit{et al.},
Mon. Not. Roy. Astron. Soc. \textbf{524}, no.2, 1885 (2023)
[arXiv:2302.02028].

\bibitem{Sanejouand:2023jkv}
Y.~H.~Sanejouand,
New Astron. \textbf{116}, 102331 (2025)
[arXiv:2312.05896].

\bibitem{Cai:2021wgv}
R.~G.~Cai \textit{et al.},
Phys. Rev. D \textbf{103}, no.12, 121302 (2021)
[arXiv:2102.02020].

\bibitem{SDSS:2014iwm}
M.~Betoule \textit{et al.},
Astron. Astrophys. \textbf{568}, A22 (2014)
[arXiv:1401.4064].

\bibitem{Wang:2015tua}
Y.~Wang and M.~Dai,
Phys. Rev. D \textbf{94}, no.8, 083521 (2016)
[arXiv:1509.02198].

\bibitem{SNLS:2011lii}
A.~Conley \textit{et al.},
Astrophys. J. Suppl. \textbf{192}, 1 (2011)
[arXiv:1104.1443].

\bibitem{Perivolaropoulos:2023iqj}
L.~Perivolaropoulos and F.~Skara,
Mon. Not. Roy. Astron. Soc. \textbf{520}, no.4, 5110 (2023)
[arXiv:2301.01024].

\bibitem{Liu:2024vlt}
Y.~Liu, H.~W.~Yu and P.~X.~Wu,
Phys. Rev. D \textbf{110}, no.2, L021304 (2024)
[arXiv:2406.02956].

\bibitem{Liddle:2007fy}
A.~R.~Liddle,
Mon. Not. Roy. Astron. Soc. \textbf{377}, L74 (2007)
[astro-ph/0701113].

\bibitem{Liddle:2009xe}
A.~R.~Liddle,
Ann. Rev. Nucl. Part. Sci. \textbf{59}, 95 (2009)
[arXiv:0903.4210].

\bibitem{Akaike:1974}
H.~Akaike,
IEEE Trans. Automatic Control \textbf{19}, 716 (1974).

\bibitem{Schwarz:1978}
G.~Schwarz, Ann. Stat. \textbf{6}, 461 (1978).

\bibitem{Perivolaropoulos:2022khd}
L.~Perivolaropoulos and F.~Skara,
Universe \textbf{8}, no.10, 502 (2022)
[arXiv:2208.11169].

\bibitem{Torrado:2020dgo}
J.~Torrado and A.~Lewis,
JCAP \textbf{2105}, 057 (2021)
[arXiv:2005.05290].

\bibitem{Cobaya}
https:$/\!/$cobaya.readthedocs.org

\bibitem{Lewis:2019xzd}
A.~Lewis,
arXiv:1910.13970 [astro-ph.IM].

\bibitem{GetDist}
https:$/\!/$getdist.readthedocs.io

\bibitem{Chen:2018dbv}
L.~Chen, Q.~G.~Huang and K.~Wang,
JCAP \textbf{1902}, 028 (2019)
[arXiv:1808.05724].

\bibitem{Fixsen:2009ug}
D.~J.~Fixsen,
Astrophys. J. \textbf{707}, 916 (2009)
[arXiv:0911.1955].

\bibitem{Kass:1995loi}
R.~E.~Kass and A.~E.~Raftery,
J. Am. Statist. Assoc. \textbf{90}, no.430, 773 (1995).

\bibitem{Kilbinger:2009by}
M.~Kilbinger {\it et al.},
Mon. Not. Roy. Astron. Soc. \textbf{405}, 2381 (2010)
[arXiv:0912.1614].

\bibitem{Weinberg:2009rd}
M.~D.~Weinberg,
arXiv:0911.1777 [astro-ph.IM].

\bibitem{Trotta:2008qt}
R.~Trotta,
Contemp. Phys. \textbf{49}, 71 (2008)
[arXiv:0803.4089].

\bibitem{Mukherjee:2017oom}
P.~Mukherjee {\it et al.},
Eur. Phys. J. Plus \textbf{134}, no.4, 147 (2019)
[arXiv:1710.02417].

\bibitem{Heavens:2017hkr}
A.~Heavens {\it et al.},
Phys. Rev. Lett. \textbf{119}, no.10, 101301 (2017)
[arXiv:1704.03467].

\bibitem{Heavens:2017afc}
A.~Heavens {\it et al.},
arXiv:1704.03472 [stat.CO].

\bibitem{MCEvidence}
https:$/\!/$github.com/yabebalFantaye/MCEvidence

\bibitem{MCEvimod}
https:$/\!/$github.com/BorisNgHL/MCEvi$_{-}$mod

\bibitem{Kenworthy:2022jdh}
W.~D.~Kenworthy {\it et al.},
Astrophys. J. \textbf{935}, no.2, 83 (2022)
[arXiv:2204.10866].

\bibitem{Marra:2021fvf}
V.~Marra and L.~Perivolaropoulos,
Phys. Rev. D \textbf{104}, no.2, L021303 (2021)
[arXiv:2102.06012].

\bibitem{Alestas:2021nmi}
G.~Alestas, I.~Antoniou and L.~Perivolaropoulos,
Universe \textbf{7}, no.10, 366 (2021)
[arXiv:2104.14481].

\bibitem{Sapone:2020wwz}
D.~Sapone, S.~Nesseris and C.~A.~P.~Bengaly,
Phys. Dark Univ. \textbf{32}, 100814 (2021)
[arXiv:2006.05461].

\bibitem{Caldwell:2005xb}
R.~R.~Caldwell {\it et al.},
Phys. Rev. D \textbf{73}, 023513 (2006)
[astro-ph/0507622].

\bibitem{Khosravi:2017hfi}
N.~Khosravi {\it et al.},
Phys. Rev. D \textbf{99}, no.10, 103526 (2019)
[arXiv:1710.09366].

\bibitem{Alestas:2021luu}
G.~Alestas {\it et al.},
Phys. Rev. D \textbf{105}, no.6, 063538 (2022)
[arXiv:2110.04336].

\bibitem{Perivolaropoulos:2022txg}
L.~Perivolaropoulos and F.~Skara,
Phys. Rev. D \textbf{106}, no.4, 043528 (2022)
[arXiv:2203.10374].

\bibitem{Wei:2021xek}
H.~Wei and Z.~X.~Yu,
JCAP \textbf{2108}, 011 (2021)
[arXiv:2103.12696].

\bibitem{Caldwell:2007yu}
R.~R.~Caldwell and A.~Stebbins,
Phys. Rev. Lett. \textbf{100}, 191302 (2008)
[arXiv:0711.3459].

\bibitem{Deng:2018yhb}
H.~K.~Deng and H.~Wei,
Phys. Rev. D \textbf{97}, no.12, 123515 (2018)
[arXiv:1804.03087].

\bibitem{Deng:2018jrp}
H.~K.~Deng and H.~Wei,
Eur. Phys. J. C \textbf{78}, no.9, 755 (2018)
[arXiv:1806.02773].

\bibitem{Carlberg:1995aq}
R.~Carlberg {\it et al.},
Astrophys. J. \textbf{462}, 32 (1996)
[astro-ph/9509034].

\bibitem{Haslbauer:2020xaa}
M.~Haslbauer, I.~Banik and P.~Kroupa,
Mon. Not. Roy. Astron. Soc. \textbf{499}, 2845 (2020)
[arXiv:2009.11292].

\bibitem{Mazurenko:2023sex}
S.~Mazurenko {\it et al.},
Mon. Not. Roy. Astron. Soc. \textbf{527}, 4388 (2024)
[arXiv:2311.17988].

\bibitem{Mazurenko:2024gwj}
S.~Mazurenko, I.~Banik and P.~Kroupa,
Mon. Not. Roy. Astron. Soc. \textbf{536}, 3232 (2025)
[arXiv:2412.12245].

\bibitem{Banik:2025dlo}
I.~Banik and V.~Kalaitzidis,
Mon. Not. Roy. Astron. Soc. \textbf{540}, no.1, 545 (2025)
[arXiv:2501.17934].

\bibitem{Haridasu:2024ask}
B.~S.~Haridasu, P.~Salucci and G.~Sharma,
Mon. Not. Roy. Astron. Soc. \textbf{532}, 2234 (2024)
[arXiv:2403.06859].

\bibitem{Aluri:2022hzs}
P.~K.~Aluri \textit{et al.},
Class. Quant. Grav. \textbf{40}, no.9, 094001 (2023)
[arXiv:2207.05765].

\bibitem{Krishnan:2021jmh}
C.~Krishnan \textit{et al.},
Phys. Rev. D \textbf{105}, no.6, 063514 (2022)
[arXiv:2106.02532].

\bibitem{McConville:2023xav}
R.~Mc Conville and E.~\'O.~Colg\'ain,
Phys. Rev. D \textbf{108}, no.12, 123533 (2023)
[arXiv:2304.02718].

\bibitem{Boubel:2024cmh}
P.~Boubel \textit{et al.},
JCAP \textbf{2503}, 066 (2025)
[arXiv:2412.14607].

\bibitem{Moffat:2016dbd}
J.~W.~Moffat,
arXiv:1608.00534 [astro-ph.CO].

\bibitem{Moffat:1991qj}
J.~W.~Moffat and D.~C.~Tatarski,
Phys. Rev. D \textbf{45}, 3512 (1992).

\bibitem{Moffat:1994qy}
J.~W.~Moffat and D.~C.~Tatarski,
Astrophys. J. \textbf{453}, 17 (1995)
[astro-ph/9407036].

\bibitem{Moffat:2025whe}
J.~Moffat,
arXiv:2502.20494 [astro-ph.CO].

\bibitem{Castello:2021uad}
S.~Castello, M.~H\"og\r{a}s and E.~M\"ortsell,
JCAP \textbf{2207}, 003 (2022)
[arXiv:2110.04226].

\bibitem{Camarena:2022iae}
D.~Camarena \textit{et al.},
Class. Quant. Grav. \textbf{39}, no.18, 184001 (2022)
[arXiv:2205.05422].

\bibitem{Montani:2024pou}
G.~Montani \textit{et al.},
Phys. Dark Univ. \textbf{48}, 101848 (2025)
[arXiv:2404.15977].

\bibitem{Montani:2024ntj}
G.~Montani, N.~Carlevaro and M.~G.~Dainotti,
Phys. Dark Univ. \textbf{48}, 101847 (2025)
[arXiv:2411.07060].

\bibitem{Dainotti:2021pqg}
M.~G.~Dainotti \textit{et al.},
Astrophys. J. \textbf{912}, no.2, 150 (2021)
[arXiv:2103.02117].

\bibitem{Schiavone:2022wvq}
T.~Schiavone \textit{et al.},
Mon. Not. Roy. Astron. Soc. \textbf{522}, no.1, L72 (2023)
[arXiv:2211.16737].

\bibitem{Silva:2023wvs}
C.~Silva,
arXiv:2312.05267 [gr-qc].

\bibitem{Vagnozzi:2023nrq}
S.~Vagnozzi,
Universe \textbf{9}, no.9, 393 (2023)
[arXiv:2308.16628].

\bibitem{Wei:2009xg}
H.~Wei,
Phys. Lett. B \textbf{682}, 98 (2009)
[arXiv:0907.2749].

\bibitem{Wei:2017mzf}
H.~Wei and D.~Z.~Xue,
Commun. Theor. Phys. \textbf{68}, no.5, 632 (2017)
[arXiv:1706.04063].

\end{thebibliography}
\end{document}